\documentclass[prd,twocolumn,showpacs,superscriptaddress,floatfix,nofootinbib]{revtex4}
\usepackage{amsfonts,amsmath,units,wasysym,epsfig,verbatim,color,subfigure,graphicx,bm,mathrsfs,lipsum}
\usepackage{amssymb}
\usepackage{multirow}

\begin{document}

\newcommand{\IUCAA}{Inter-University Centre for Astronomy and
  Astrophysics, Post Bag 4, Ganeshkhind, Pune 411 007, India}

\newcommand{\WSU}{$^*$Department of Physics \& Astronomy, Washington State University,
1245 Webster, Pullman, WA 99164-2814, U.S.A \\}

\title{Probing the nature of central objects in extreme-mass-ratio inspirals with gravitational waves}
\author{Sayak Datta} 
\affiliation{\IUCAA}

\author{Sukanta Bose}
\affiliation{\IUCAA}
\affiliation{\WSU}

\date{\today}

\begin{abstract}
We extend the work of Ryan~\cite{Ryan,Ryan2} on mapping the spacetime of the central object of an extreme mass-ratio inspiral (EMRI) by using gravitational waves (GWs) emitted by the system, which may be observed in future missions such as LISA. Whether the central object is a black hole or not can be probed by observing the phasing of these waves, which carry information about its mass and spin multipole moments. We go beyond the phase terms found by Ryan, which were obtained in the quadrupolar approximation of the point-particle limit, and derive terms up to the fifth post-Newtonian (PN) order. Since corrections due to horizon absorption (i.e., if the central object is a black hole) and tidal heating appear by that order, at 2.5PN and 5PN, respectively, we include them here. Corrections due to the motion of the central object, which was addressed only partially by Ryan, are included as well. Additionally, we obtain the contribution of the higher order radiative multipole moments. For the tidal interaction, our results have been derived in the approximation of the Newtonian tidal field. Therefore, in the potential for tidal field only the contribution due to the mass of the central object has been included as well. Using these results we argue that it might be possible for LISA to probe if the central object in an EMRI has a horizon or not. We also discuss how our results can be used to test the No-hair theorem from the inspiral phase of such systems.

\end{abstract}

\maketitle

\section{Introduction}
\label{sec:Intro}

The direct observation of gravitational waves (GWs)\cite{LIGO detection 1,LIGO detection2,LIGO detection3,LIGO detection4} has opened up a new vista onto the universe. The Laser Interferometer Space Antenna~(LISA)~\cite{LISA 1,LISA 2,LISA 3,LISA 4,LISA 5}, which is likely to be launched in the early-to-mid 2030s, is expected to increase the variety of objects that will be observed in GWs. If the history of science is any indication, then it is not inconceivable that we will eventually observe GW systems that we did not think of before. Characterizing any GW source, however, benefits from the ability to map its space-time. A formalism that allows one to do so can, therefore, be useful. By making a few assumptions, Ryan~\cite{Ryan,Ryan2} showed that the waves emitted by a small compact body orbiting a much more massive compact object carry information about the Geroch-Hansen multipole moments of the latter~\cite{Geroch,Hansen}. These moments characterize the heavier object's vacuum spacetime geometry in what is termed as an extreme mass-ratio inspiral (EMRI) system. 

As an example, if the massive body is a black hole then the ``No-hair" theorem~\cite{no hair israel,no hair israel2,no hair wald,no hair carter,no hair Robinson} states that its exterior metric found by solving the Einstein-Maxwell equations of gravitation and electromagnetism in general relativity will be dependent only on mass, electric charge and angular momentum. Owing to this theorem, we expect the moments of black hole solutions to depend only on these parameters. 
For this reason, measuring the multipole moments from observations will help probe the validity of the No-hair theorem or constrain it. To pursue this goal of mapping the spacetime of the heavy object in the approximate center of EMRIs we will follow the formalism constructed by Ryan~\cite{Ryan}. In Sec.~\ref{sec:obs}, we will discuss it briefly since it will be useful in describing the properties of the central object and, therefore, test the theorem.

Since Ryan's work in the late 1990s, it was realized that this formalism has an important limitation in the sense that the orbits studied there are not realistic for EMRIs. The orbits studied there were equatorial and circular, while the realistic orbits are expected to be eccentric and nonequatorial. Effects of self-force, tidal deformation and absorption of GW by the central object were also ignored. These drawbacks were partially addressed in Refs. \cite{Barack 2004, Glampedakis, ecentric orbit 1, ecentric orbit 2, ecentric orbit 3, ecentric orbit 4, Hughes 2001, Taracchini 2013, Pound 2008, Meent 2018, absorption, Tagoshi, Li}, which helped in the development of better waveforms. Owing to the system's extreme mass ratio, a way to generate realistic EMRI waveforms is to use
black hole perturbation theory, governed by Teukolsky equation. Such calculations are computationally very expensive~\cite{Barack 2009, self-force}. To tackle this issue approximate waveforms, known as {\em kludge} waveforms, have been constructed~\cite{Babak 2007, Chua 2017}. These latter waveforms are not perfect either since they do not account for self-force. Most accurate EMRI waveforms can still be produced, however, with numerical evolution of the orbit. Unfortunately, they are computationally expensive to produce. This is a major reason why many such studies have resorted to using the kludges, with the latest such work being the one in Ref.~\cite{Gair 2017}. 

In this work, we  study the viability of searching for the existence of horizon and tidal deformablity, and testing for the No-hair theorem, with LISA. Since the use of numerical waveforms for this purpose is computationally prohibitive at this stage, such an analysis should ideally be carried out at least with kludge waveforms, after extending them to account for the horizon term, the tidal deformability parameter, etc. However, before we do so, in a future work, we first conduct a viability test here for the measurement of 
the horizon term, the tidal deformability parameter, and the leading mass and spin multipole moments of the central objects in EMRIs, for it is not clear with any of the aforementioned waveform families how accurately they will be measurable in LISA. This is a first step in that regard, which we aim to improve in subsequent work with better waveforms.

One of the effects that Ryan's formalism does not account for is matter tides. If the central object is not a black hole, its effect on the GW phasing can be non-negligible~\cite{Isoyama1,Isoyama2}.
Li et al. \cite{Li} have included this effect in an extension of Ryan's formalism. A second effect not considered in that formalism is the absorption of gravitational waves by the central object if it happens to have a horizon. We study both these effects here. It will be shown how these can be used to figure out whether the central object has a horizon and, consequently, to test the No-hair theorem. 

A third important contribution absent in Ryan's work is the effect of radiation reaction. When the two objects in an EMRI are far apart, the GW luminosity is small, and this effect is not very prominent. Over time, however, it has a cumulative effect on the phase of the gravitational wave emitted by it. In this work, we did not account for this correction either. This is along the lines of the other simplifying assumptions we have made, e.g., by ignoring eccentricity, precession, etc., which we expect to be present in realistic EMRIs. In that sense,  the present work should be considered as a first step towards the more complex goal of assessing how precisely parameters of astrophysically realistic EMRIs can be measured in LISA. In spite of this assumption, we feel that the current work serves a useful role since it provides some understanding for the first time of the observability of each of the first few multipoles. And the other important aspect is the formulation of the data analysis framework involving the horizon parameter, which brings us closer to probe the existence or absence of a horizon.

\par 

In Sec.~\ref{sec:ryan} we briefly discuss Ryan's formalism. 
There we also describe tidal effects and give the expressions for the observables related to the vacuum multipole moments. 
We calculate the complete luminosity up to tenth power in the velocity of the orbiting companion $v$ under the point particle~(PP) approximation. This is followed in Secs.~\ref{sec:tides}-\ref{sec:Luminosity absorbed inside the horizon} by a presentation of the luminosity absorbed by the horizon in the case where the central object has one. Next, in Sec.~\ref{sec:mass dep. part} we discuss that contribution to the luminosity that depends only on the masses. In Sec.~\ref{sec:central motion} we discuss the terms in the luminosity that arise due to the central object's motion. Then in Sec.~\ref{sec:beyond ryan}  after giving the complete expression of the phase evolution, we discuss how it can be used to probe for a horizon and also test the No-hair theorem. In Sec.~\ref{sec:paramestim}, we present estimates of some parameter errors for a exploratory set of EMRIs. We use the Fisher information matrix for this purpose. Such a method has been used in the past for estimating a few of the parameters of EMRIs in LISA (e.g., sky position, total mass, mass ratio, eccentricity, and the spin of the central object~\cite{Gair 2017, Barack 2004,Seoane 2013,Seoane 2017,Babak 2014,Seoane 2015}). We extend that list to include additional parameters, such as a few of the low-order mass and spin moments, the horizon term and the tidal deformability parameter of the central object. We note that the Fisher estimation method has well known limitations: Importantly, our error estimates should be considered only for loud signals and under the simplifying assumptions listed above. We leave more sophisticated parameter estimation studies using Bayesian methods for the future when some of the assumptions about the system can be dropped in order to make the system more realistic.

Throughout this work we have used $G=1=c$.

\section{EMRI Observables}
\label{sec:obs}

In Ref.~\cite{Ryan} Ryan showed how  the multipole moments of the central object can be extracted from the gravitational waves emitted by an 
EMRI. 
He showed that certain functions, discussed below, are ``good" observables for this  purpose.

One such quantity is the gravitational-wave spectrum ${\Delta}E(f)$, which is defined as
\begin{align}
\label{spectrum}\Delta E \equiv&  - \Omega \frac{dE_{\rm source}}{d\Omega}\,,
\end{align}
where $E_{\rm source}$ is the energy of the binary system and $\Omega$ is the orbital frequency defined as,

\begin{equation}
\Omega = \frac{d\phi}{dt} = \frac{d\phi}{d\tau}\frac{d\tau}{dt},
\end{equation}
where, $\tau$ is the proper time along the geodesic.

Another observable quantity is the phase evolution,
which we will define as the rate of change of the primary wave frequency with time. 
The following dimensionless wave observable quantifies it~\cite{Ryan}:
\begin{align}
\label{N}\Delta N(f) \equiv& \frac{f^2}{df/dt} = \frac{f \Delta E(f)}{-dE_{\rm Total}/dt}\,,
\end{align}
where $f$ is the GW frequency, and is related to the orbital frequency as $f = 2\Omega$. Moreover, $-dE_{\rm Total}/dt$ is the total emitted luminosity from the system in the form of GWs; the minus sign represents this {\it loss} of energy from the orbit.  In addition to $\Delta N$, we will be interested in studying the waveform phase $(\psi)$, which is related to the former as follows \cite{Flanagan,phase expression}:
\begin{equation}
\begin{split}
\psi(f) = 2\pi f t_c - 2\phi_c-\frac{\pi}{4} + 6\int^v_{v_i} d\bar{v} (v^3-\bar{v}^3) \frac{\pi\Delta N}{\bar{v}^4}\,,
\end{split}
\end{equation}
where $v$ is the orbital velocity of the smaller body, $\bar{v}$ is an integral variable for velocity, $v_i$ is some initial reference point for velocity and $M$ is the total mass of the system. $t_c = t(v_i)$ and $\phi_c = \phi(v_i)$ \cite{Poisson 1995}.

For orbits that are slightly elliptical and slightly inclined to the equatorial plane, there are two other observables in the form of precession frequencies. Owing to the near-axisymmetry of an EMRI, we employ the cylindrical co-ordinate system to describe them, with $\rho$ as the radial co-ordinate and $z$ as the axial co-ordinate. Then one of the frequencies is related to the rate at which $\rho$ changes and the other to the rate at which $z$ changes~\cite{Ryan}:
\begin{widetext}
\begin{equation}
\begin{split}
\Omega_{\alpha}=&\,\, {\Omega}-\bigg\{-{\frac{g^{\alpha\alpha}}{2}}\big[(g_{tt}+{\Omega}g_{t\phi})^2\bigg(\frac{g_{\phi\phi}}{{\rho}^2}\bigg)_{,{\alpha\alpha}}-2(g_{tt}+{\Omega}g_{t\phi})(g_{t\phi}+{\Omega}g_{\phi\phi})(\frac{g_{t\phi}}{{\rho}^2})_{,\alpha\alpha}+(g_{t\phi}+{\Omega}g_{\phi\phi})^2(\frac{g_{tt}}{{\rho}^2})_{,\alpha\alpha}\big]\bigg\}^{\frac{1}{2}}\,,
\end{split}
\end{equation}
\end{widetext}
where $\alpha = \rho$ and $z$, respectively, for the two cases, and there is no sum over $\alpha$ on the right-hand side (RHS).

Since we will include the effect of tidal interaction, we will examine, in particular, how any of these observables gets modified by its presence and how one can extract  multipole information from them.

The precession frequencies depend on the possibly complex orbit and its orientation. Tides are related to the deformation of the inspiraling bodies and, thus, to first order will not affect the precession frequencies ~\cite{Li}. Then remains the phase evolution. As there are tidal corrections to the luminosity, we can anticipate that $\Delta N$ will be affected by tidal deformation, when present, even if at higher orders of $v$.
 
 \section{Ryan's Formalism}
\label{sec:ryan}

Since our aim is to generalize the multipole formalism for EMRIs in Ref.~\cite{Ryan}, we begin by briefly summarizing it.

\label{Assumptions}\subsection{Assumptions}

The assumptions of the formalism are:
\par
(i) The central body of the system has a vacuum, external gravitational field that is stationary, axisymmetric, reflection symmetric across the equatorial plane of the central object, and asymptotically flat (SAVAR). In the $(t,{\rho},{\phi},z)$ co-ordinate system the metric takes the form~\cite{Ryan},
\begin{align}
\label{metric}ds^2 = -F(dt - {\omega}{d\phi})^2 + {\frac{1}{F}}[e^{2\gamma}({d\rho}^2+{dz}^2)+{\rho}^2{d\phi}^2]\,,
\end{align}
where $F$ and $\omega$ can, in general, depend on $\rho$ and $z$.
\par
(ii) The companion compact object of mass $m$ inspirals around the central much heavier compact object of mass $M \gg m$.
Consequently, small perturbations of the central object's vacuum metric will not induce any significant change. Owing to this assumption it is possible to treat the inspiraling object as a ``test particle", which has an orbit evolving slowly and adiabatically from one geodesic orbit to another. So, on the time scale of the orbital period it can be approximated as a geodesic.\\
\par
(iii) The geodesic orbits through which the inspiral evolves are almost circular. In general, they can be slightly elliptical and will lie mostly in the equatorial plane.\\
\par
(iv) The central object does not absorb any energy so all the energy is emitted to infinity. (Ryan did not account for the absorption by a horizon that the central object may have.) Also, tidal effect has been neglected completely. 

 We will try to relax these assumptions as much as possible in later sections.

The parameter space of EMRIs, however, contains additional parameters that we will neglect for simplicity \cite{Barack 2004}. Another crucial point to note is that the event rate of high-eccentricity EMRIs is much larger than that of low-eccentricity ones~\cite{ecentric orbit 1, ecentric orbit 2, ecentric orbit 3, ecentric orbit 4}. High eccentricity EMRIs spend enough cycles inside the band of eLISA to be detectable~\cite{ecentric orbit 1, ecentric orbit 4}. Therefore circular orbit EMRIs are not realistic. Since the velocity of the small mass can be highly relativistic, PN expansion is not adequate either. We hope to return to those aspects in a later work, and limit our scope here to discuss for the simple EMRI systems absorption by the horizon and the signature of the multipole moments in GWs.
For this reason, the results on parameter estimates obtained in the present work can be considered as ``indicative".

\subsection{Procedure and results}
\label{subsec:Procedure and results}

Owing to the assumptions described in the previous section, the space-time mapping problem becomes easier to address. Assumption (i) limits the metric around the massive body to be a SAVAR metric. Due to the symmetries, the metric is independent of $t$ and $\phi$, thereby, implying the existence of two conserved quantities, namely, energy $E$ and angular momentum $L_{z}$~\cite{Ryan}:
\begin{align}
\label{E}\frac{E}{m} = - g_{tt}\bigg(\frac{dt}{d\tau}\bigg) - g_{t\phi}\bigg(\frac{d\phi}{d\tau}\bigg)\,,
\end{align}
\begin{align}
\label{L}\frac{L_{z}}{m} = g_{t\phi}\bigg(\frac{dt}{d\tau}\bigg) + g_{\phi\phi}\bigg(\frac{d\phi}{d\tau}\bigg)\,,
\end{align}
where $m$ is the mass of the lighter orbiting object, and $\tau$ is the proper time along its geodesic. $M$ will denote the mass of the heavy central object, 
The rotational frequency $\Omega$ of a circular orbit can be expressed as,
\begin{align}
\label{Circular frequency}\Omega = \frac{d\phi}{dt} = \frac{-g_{t{\phi},\rho} + \sqrt{(g_{t{\phi},\rho})^2 - g_{tt,\rho}g_{\phi\phi{,}\rho}}}{g_{\phi\phi{,}\rho}}.
\end{align}
From the normalization equation of the four velocity, and using Eqs.~(\ref{L}) and (\ref{E}) one finds,
\begin{align}
\label{time velocity}\frac{dt}{d\tau} = \frac{1}{\sqrt[]{ -g_{tt}- {\Omega}^2 g_{\phi\phi} - 2{\Omega}g_{t\phi}}}.
\end{align}
Substituting the last two expressions in the energy and angular momentum equations, Eqs.~(\ref{E}) and (\ref{L}), respectively, one obtains
\begin{align}
\label{E2}\frac{E}{m} = \frac{- g_{tt} -{\Omega}g_{t\phi}}{\sqrt{- g_{tt} - 2g_{t\phi}{\Omega}- g_{\phi\phi}{\Omega}^2}}\,,
\end{align}
\begin{align}\
\label{L2}\frac{L_{z}}{m} = \frac{g_{t\phi} +{\Omega}g_{\phi\phi}}{\sqrt{- g_{tt} - 2g_{t\phi}{\Omega}- g_{\phi\phi}{\Omega}^2}}.
\end{align}
Using the above equations, the Ernst potential formalism and the results from Ref.~\cite{fodor}, Ryan found the expressions of the aforementioned wave observables to be~\cite{Ryan}
\begin{widetext}
\begin{equation}
\begin{split}
\label{delta E}\frac{\Delta E }{m}=&\frac{v^2}{3}-\frac{v^4}{2} +\frac{20 S_1 v^5}{9 M^2}+ \left(\frac{M_2}{M^3}-\frac{27}{8}\right) v^6+\frac{28 S_1 v^7}{3 M^2}+v^8 \left(\frac{80 S_1^2}{27 M^4}+\frac{70 M_2}{9 M^3}-\frac{225}{16}\right)+...\,,
\end{split}
\end{equation}
\begin{align}
\label{frequency 1}\frac{\Omega_{\rho}}{\Omega} &= 3v^2 - 4\frac{S_1}{M^2} v^3 + \bigg(\frac{9}{2} - \frac{3}{2} \frac{M_2}{M^3}\bigg)v^4 - 10\frac{S_1}{M^2} v^5 + \bigg(\frac{27}{2} - 2\frac{S_1^2}{M^4} - \frac{21}{2}\frac{M_2}{M^3}\bigg)v^6+ ...\,,
\end{align}
\begin{align}
\label{frequency 2}\frac{\Omega_z}{\Omega} &= 2\frac{S_1}{M^2} v^3 + \frac{3}{2}\frac{M_2}{M^3}v^4 + \bigg(7\frac{S_1^2}{M^4} + 3\frac{M_2}{M^3}\bigg)v^6 + \bigg(11\frac{S_1 M_2}{M^5} - 6\frac{S_3}{M^4}\bigg)v^7+ ...\,,
\end{align}
\end{widetext}
where $v$ is the orbital velocity of the lighter companion about the center of mass and $S_l$ and $M_l$ are respectively the current and the mass multipole moments defined by Hansen~\cite{Hansen} and Geroch~\cite{Geroch}. The symbol $M$ denotes a mass parameter. However, confusion with mass multipole moment $M_l$ can be avoided by noting that the latter has a subscript but the former does not.

This is how Ryan achieved the goal of expressing the observables in terms of the central object's multipole moments. To calculate the phase evolution it is important to know the luminosity of the system, which we 
discuss later.

\section{Matter tides}
\label{sec:tides}

\subsection{Result of tidal effect on a compact star}

In the last section we discussed how Ryan derived the expressions for certain GW observables, under a set of assumptions. But due to assumption (iv), tidal interaction was neglected in his formalism. In this section we discuss how it can be included and what changes it brings about in the observables.
\par
Let us take $m_1$ and $m_2$ to be the masses of the inspiraling compact objects. Also, let $\Omega$ be the orbital angular frequency and $\mu_r$ the reduced mass of the system. Then 
\begin{align}
\eta = \frac{m_1 m_2}{M_{T}^2} = \frac{\mu_r}{M_T},\,\,\,\,M_T = m_1 + m_2,
\end{align}
are the symmetrized mass-ratio and the total mass of the binary, respectively.

Flanagan and Hinderer~\cite{Flanagan} calculated the energy in gravitational waves associated with tidal effects and the contribution to the rate of change of energy due to them.  To set the stage for our calculations, we begin by discussing their results first. They took the effective action of the inspiraling system and an associated quadrupole moment coupled with orbits through a tidal field. From its solution, they calculated the induced quadrupole moment. To more precisely point out the physical arguments it is useful to start with the action they considered. Suppose the relative separation of the two objects is $x^i = (\rho \cos\Phi , \rho \sin\Phi , 0) = \rho \,n^i$. Let $Q^{(n)}_{1ij}$ be the quadrupolar deformation of the first object caused by the tidal field $\mathscr{E}_{2ij} = -m_2 \partial_i \partial_j(1/\rho)$ of the second object. Here we limit ourselves to the 
$l = 2$ order, and with $n$ radial nodes. Then, $Q_{1ij} = \sum_n Q^{(n)}_{1ij}$ and the tidal deformability of the first object, $\lambda_1 = \sum_n \lambda_{1,n}$.
Under these conditions the action for the system is~\cite{Flanagan}

\begin{widetext}
\begin{equation}
\begin{split}
S = \int dt \bigg[\frac{1}{2}\mu_r \dot{\rho}^2 + \frac{1}{2}\mu_r {\rho}^2 \dot{\Phi}^2 + \frac{M_T\mu_r}{\rho}\bigg] - \{\frac{1}{2}\int dt \, Q_{1ij} \,\mathscr{E}_{2ij} - \sum_n \int dt \frac{1}{4\lambda_{1,n} \omega_n^2} \bigg[\dot{Q}^{(n)}_{1ij} \dot{Q}^{(n)}_{1ij} - \omega_n^2Q^{(n)}_{1ij} Q^{(n)}_{1ij}\bigg] + 1\leftrightarrow2 \,\}.
\end{split}
\end{equation} 
\end{widetext}
If the Burke-Thorne GW dissipation contribution~\cite{Maggiore} is ignored then the ensuing equations of motion for the first object are
\begin{align}
\label{orbit eqn}\ddot{x}^i + \frac{M_T}{\rho^2}n^i = \frac{m_2}{2\mu_r}Q_{1jk}\partial_i\partial_j\partial_k\frac{1}{\rho},\\
\label{quadrupole eqn}\ddot{Q}^{(n)}_{1ij} + \omega_n^2 Q^{(n)}_{1ij} = m_2\lambda_{1,n}~\omega_n^2\partial_i\partial_j\frac{1}{\rho}\,,
\end{align}
where $x^i$ are its spatial coordinates. These equations have equilibrium solutions with $\rho$ as constant and $\Phi = \Phi_0 + \omega t$. The second object's equations of motion can be calculated similarly.  

By accounting for the contributions from both of the bodies, Flanagan et al. obtained the expression for the orbital radius, the energy of the binary and the GW luminosity. The results we will use for the energy and luminosity are~\cite{Hinderer}:
\begin{align}
E_{\rm Tidal}(v) &= {\frac{9}{2}}\frac{\eta v^{12}}{M^4{_T}} \bigg[{\frac{m_1}{m_2}}{\lambda}_{2}+ 1\leftrightarrow2\bigg]\\
\frac{dE}{dt}\bigg{|}_{\rm Tidal}(v) &=- {\frac{32}{5}}{\eta}^2\frac{v^{20}}{M_{T}^5}6\bigg[\frac{m_1 + 3m_2}{m_1}{\lambda}_{1}+1\leftrightarrow 2\bigg],
\end{align}
where $v^2 = \{\Omega(m_1+m_2)\}^{2/3}$.

For our purpose we will denote the parameters of the more massive, central object with the index $M$ and those of the second object with the index $m$. In other words, replacing the indices 1 and 2 with $M$ and $m$, respectively, the above equations become:
\begin{align}
\label{tidal energy}
E_{\rm Tidal}(v) &= {\frac{9}{2}}\frac{\eta v^{12}}{M^4{_T}} \bigg[{\frac{M}{m}}{\lambda}_{m}+{\frac{m}{M}}{\lambda}_{M}\bigg],
\end{align}
\begin{equation}
\begin{split}
\label{tidal energy rate}\frac{dE}{dt}\bigg{|}_{\rm Tidal}(v) &=- {\frac{32}{5}}{\eta}^2\frac{v^{20}}{M_{T}^5}6\bigg[\frac{m + 3M}{m}{\lambda}_{m}\\
&+\frac{M + 3m}{M}{\lambda}_{M}\bigg]\\
&=-\frac{32}{5}\frac{m^2}{M^2} Av^{20},
\end{split}
\end{equation}
where in the limit of the extreme-mass ratio, $v^2 = (M \Omega)^{\frac{2}{3}}$, $A = \frac{M^2}{m^2}\frac{6{\eta}^2}{M_T^5}\big\{\frac{m + 3M}{m}{\lambda}_{m}+\frac{M + 3m}{M}{\lambda}_{M}\big\}$ and $M_T = M$. It straightforwardly follows from the above that 
\begin{equation}
\begin{split}
\label{Tidal Delta E}\Delta E &=-18\frac{\eta v^{12}}{M^4} \bigg[{\frac{M}{m}}{\lambda}_{m}+{\frac{m}{M}}{\lambda}_{M}\bigg]\\ 
&= Xv^{12},
\end{split}
\end{equation}
 where, $X=-18\frac{\eta }{M^4} \big\{{\frac{M}{m}}{\lambda}_{m}+{\frac{m}{M}}{\lambda}_{M}\big\}$. When applying the above result, one must note that  the tidal Love number of a black hole is zero~\cite{TLN of BH 1,TLN of BH 2,tidal perturbation of sc bh,TLN of BH 3,TLN of BH 4,TLN of BH 5}. We replace $\lambda_M$ and $\lambda_m$ with $\lambda_M/M^5 = \Lambda_M$ and $\lambda_m/m^5 = \Lambda_m$, where $\Lambda_m$ and $\Lambda_M$ are the dimensionless tidal deformability.

Higher order contributions due to tidal interactions, including those beyond Ref.~\cite{Flanagan}, have been calculated by Damour et al.~\cite{Damour}, but for the mass-dependent tidal field alone. However, the tidal field depends on the multipolar structure of the source. Contributions from higher order multipoles were not considered in their work. Since the tidal corrections obtained in Flanagan et al.~\cite{Flanagan} are at the lowest order, it is consistent  to use their results for our multipolar study.
\newline

\subsection{Black hole as central massive object}

In the previous section we discussed how the tidal perturbation of a compact star contributes to the GW emission of a binary. Since a black hole has a vanishing tidal Love number the tidal terms there will not contribute to GWs emitted by an EMRI constituted of black holes. But there could still be induced quadrupole moment in the case of a black hole~\cite{tidal perturbation of sc bh}. We expect that the tidal distortion of the central black hole, due to its companion, will contribute at very high orders in the GWs.
To justify our point, here we look into the tidal contribution if the central object is a Schwarzschild black hole. For this only the Newtonian tidal interaction has been considered. The result has been derived by Li et al.~\cite{tidal perturbation of sc bh},~\cite{Li}:  

\begin{align}
&I_{ij}^{\rm induced} =  \frac{32}{45} M^6 \dot{\mathscr{E}}_{ij}^{\rm external},\\
&\mathscr{E}_{ij}^{\rm external} =  \frac{m}{\rho^3} (\delta_{ij} - n_i n_j),\,\,\,\, {\rm where}\\
&n_1 =  \cos(\Omega t),  \,\,\,\,\, n_2 =  \sin(\Omega t),  \,\,\,\,\, n_3 = 0.
\end{align}
Using this induced quadrupole  moment $(I^{\rm induced}_{ij})$ in the multipole formula of radiation luminosity, Eq.~(\ref{time avg}), we find
\begin{align}
-\frac{dE}{dt} = \frac{131072}{10125} \frac{m^2 M^4}{\rho^6} v^{24}.
\end{align}
Since $\frac{1}{\rho^6} = \frac{v^{12}}{M^6}
$, the luminosity simplifies to

\begin{align}
-\frac{dE}{dt} = \frac{131072}{10125} \frac{m^2}{M^2} v^{36}.
\end{align}
So, we can see that the tidal distortion of a black hole due to the Newtonian potential of the companion occurs at higher order. For that reason we do not consider this tidal distortion any further in this paper.

\section{Complete Point Particle result through $v^{10}$}
\label{sec:PP result}

In Sec.~\ref{sec:tides} we discussed the tidal contribution and how it can be included in Ryan's formalism. Here we will discuss how we can find the expression for the phase evolution $\Delta N$. From Eq.~(\ref{N}) we can see that to find $\Delta N$ we first obtain the GW luminosity emitted by the inspiraling system. The gravitational wave luminosity can be determined by calculating symmetric trace free (STF) moments ~\cite{Thorne} of the system. For the central body's Geroch-Hansen moments we use $M_l$ and $S_l$, and for the radiative moments of the complete system we use $I_L$ and $J_L$. Here $L$ is a shorthand for $b_1...b_l$, where $b_k$ is a spatial index, and $k$ and $l$ are positive integers.

In terms of these moments the radiated luminosity becomes~\cite{Thorne},
\begin{widetext}
\begin{equation}
\begin{split}
\label{time avg}-\frac{dE}{dt} = &{\sum_{l=2}^{\infty}}\frac{(l+1)(l+2)}{l(l-1)}\frac{1}{l!(2l+1)!!}\langle{I_{L}^{(l+1)}I_{L}^{(l+1)}}\rangle + {\sum_{l=2}^{\infty}}\frac{4l(l+2)}{(l-1)}\frac{1}{(l+1)!(2l+1)!!}\langle{J_{L}^{(l+1)}J_{L}^{(l+1)}}\rangle \,,
\end{split}
\end{equation}
\end{widetext}
where the angular brackets indicate average over time and the parenthetic number in the superscript of a quantity denotes the number of times its time-derivative is taken, before the averaging.
In this notation the moments of the whole system are~\cite{Thorne},
\begin{align}
I_{L}(t) &= \bigg[\int {d^3}y {\tilde{\rho}}(y,t)y_L\bigg]^{\rm{STF}}\\
J_{L}(t) &= \bigg[\int {d^3}y {\tilde{\rho}}(y,t)y_{L-1}{\epsilon}_{{b_l}km}y_{k}u_{m}\bigg]^{\rm{STF}}\,,
\end{align}
where $\tilde{\rho}$ is the mass density of the system and $y_L = y_{b_1}y_{b_2}...y_{b_l}$, with $y_{b_l}$ being spatial coordinate. The leading order contribution comes from mass quadrupole radiative moment $I_{ij}$. This loss of luminosity can be written as,
\begin{align}
\label{massquadrupole}-\frac{dE}{dt}\bigg{|}_{I_{ij}} = \frac{32}{5}{m}^2{\rho}^4{\Omega}^6\,.
\end{align}
We consider the contribution in luminosity due to higher order radiative moments too. The results are:
\begin{align}
\label{massoctopole}-\frac{dE}{dt}\bigg{|}_{I_{ijk}} = \frac{2734}{315}{m}^2{\rho}^6{\Omega}^8\,,
\end{align}
\begin{align}
\label{currentquadrupole1}-\frac{dE}{dt}\bigg{|}_{J_{ij}} = \frac{8}{45}{m}^2{\rho}^6{\Omega}^8\,,
\end{align}
\begin{align}
\label{currentquadrupole2}-\frac{dE}{dt}\bigg{|}_{I_{ijkl}} = \frac{57376}{3969}{m}^2{\rho}^8{\Omega}^{10}\,,
\end{align}
\begin{align}
\label{currentquadrupole3}-\frac{dE}{dt}\bigg{|}_{I_{ijklm}} = \frac{4010276}{155925}{m}^2{\rho}^{10}{\Omega}^{12}\,,
\end{align}
\begin{align}
\label{currentquadrupole4}-\frac{dE}{dt}\bigg{|}_{J_{ijk}} = \frac{32}{63}{m}^2{\rho}^8{\Omega}^{10}\,,
\end{align}
\begin{align}
\label{currentquadrupole5}-\frac{dE}{dt}\bigg{|}_{J_{ijkl}} = \frac{11482}{11025}{m}^2{\rho}^{10}{\Omega}^{12}\,.
\end{align}
We know that as time evolves the system spirals in and its rotation frequency changes. Therefore, the change in that frequency should be related to the change in the orbital radius. The expression for the evolving radius, after accounting for each Geroch-Hansen multipole moment, was obtained by Ryan~\cite{Ryan}:
\begin{widetext}
\begin{align}
\label{radius mass multipole}\rho =& Mv^{-2} \bigg(1 + {\sum_{l=2,4...}}\frac{(-1)^{l/2} (l+1)!!\,\, M_{l}\,\, v^{2l}}{3\,\,\,\,l!!\,\, M^{l+1}}- {\sum_{l=1,3,...}}\frac{2(-1)^{(l-1)/2}\,\, l!!\,\, S_{l}\,\, v^{2l+1}}{3\,\,\,\,(l-1)!!\,\, M^{l+1}}\bigg).
\end{align}
\end{widetext}
Using it in Eq.~(\ref{massquadrupole}) Ryan~\cite{Ryan} found,
\begin{widetext}
\begin{equation}
\begin{split}
-\frac{dE}{dt}\bigg{|}_{I_{ij}} = \frac{32}{5}\bigg(\frac{m}{M}\bigg)^{2} v^{10}\bigg(1 &+ {\sum_{l=2,4...}}\frac{(-1)^{l/2}\,\, (l+1)!!\,\, M_{l}\,\, v^{2l}}{3\,\,\,\,l!!\,\,M^{l+1}} - {\sum_{l=1,3,...}}\frac{2(-1)^{(l-1)/2}\,\, l!!\,\, S_{l}\,\, v^{2l+1}}{3\,\,\,\,(l-1)!!\,\, M^{l+1}}\bigg)^4.
\end{split}
\end{equation}
\end{widetext}
He further mentioned that to test the No-hair theorem it is enough to know the series up to $v^4$, while retaining only the dominant contribution from each multipole moment. But in our case, unlike Ryan, we are considering the effect of tidal distortion as well as absorption by the central object. As we have seen, the tidal contribution comes in at an order as high as $v^{10}$. Thus, to separate out the tidal effect completely and still test the No-hair theorem, one needs knowledge of terms completely up to $v^{10}$. 
The result we find is:
\begin{widetext}
\begin{equation}
\label{quadrupolar luminosity with H.O.}
\begin{split}
-\frac{dE}{dt}\bigg{|}_{I_{ij}} = \frac{32}{5}\bigg(\frac{m}{M}\bigg)^{2} v^{10}\bigg(1 &+ {\sum_{l=2,4...}}\frac{4(-1)^{l/2}\,\, (l+1)!!\,\, M_{l}\,\, v^{2l}}{3\,\,\,\,l!!\,\,M^{l+1}} - {\sum_{l=1,3,...}}\frac{8(-1)^{(l-1)/2}\,\, l!!\,\, S_{l}\,\, v^{2l+1}}{3\,\,\,\,(l-1)!!\,\, M^{l+1}} + {\rm H.O.}\bigg)\,,
\end{split}
\end{equation}
\end{widetext}
where H.O. represents higher order terms of the binomial expansion of $\rho$ in Eq.~(\ref{quadrupolar luminosity with H.O.}). As we only need terms up to the tenth power in $v$, we will only take those pieces of H.O. that contribute up to that order; to that extent we find:
\begin{widetext}
\begin{equation}
\begin{split}
{\rm H.O.} =& +\frac{8 S_1^2 v^6}{3 M^4}+\frac{4 M_2 S_1 v^7}{M^5}+\frac{3 M_2^2 v^8}{2 M^6} -\frac{32 S_1^3 v^9}{27 M^6}-v^{10} \left(\frac{8 M_2 S_1^2}{3 M^7}+\frac{8 S_3 S_1}{M^6}\right).
\end{split}
\end{equation}
\end{widetext}
We also calculated the contributions from $I_{ijk}$, $J_{ij}$, $I_{ijkl}$, $J_{ijk}$, $I_{ijklm}$ and $J_{ijkl}$.
The ones from $I_{ijk}$ and $J_{ij}$ are
\begin{widetext}
\begin{equation}
\begin{split}
-\frac{dE}{dt}\bigg{|}_{I_{ijk} \& J_{ij}} =& \frac{62}{7} \bigg(\frac{m}{M}\bigg)^{2} v^{12}\bigg(1 + {\sum_{l=2,4...}}\frac{(-1)^{l/2} (l+1)!! M_{l} v^{2l}}{3\,\,\,\,l!!M^{l+1}}- {\sum_{l=1,3,...}}\frac{2(-1)^{(l-1)/2} l!! S_{l} v^{2l+1}}{3\,\,\,\,(l-1)!! M^{l+1}}\bigg)^6\\
=& \frac{62}{7} \bigg(\frac{m}{M}\bigg)^{2} v^{12}\bigg[1-\frac{4 S_1 v^3}{M^2}-\frac{3 M_2 v^4}{M^3}+\frac{20 S_1^2 v^6}{3 M^4} +v^7 \left(\frac{10 M_2 S_1}{M^5}+\frac{6 S_3}{M^4}\right)+\left(\frac{15 M_2^2}{4 M^6}+\frac{15 M_4}{4 M^5}\right)
   v^8\bigg].
\end{split}
\end{equation}
Those from $I_{ijkl}$ and $J_{ijk}$ are
\begin{equation}
-\frac{dE}{dt}\bigg{|}_{I_{ijkl} \& J_{ijk}} = \frac{59392}{3969} \bigg(\frac{m}{M}\bigg)^{2} v^{14}\bigg(\frac{112 S_1^2 v^6}{9 M^4}-\frac{4 M_2 v^4}{M^3}-\frac{16 S_1 v^3}{3 M^2}+1\bigg).
\end{equation}
Finally, the ones from $I_{jklmn}$ and $J_{ijkl}$ are:
\begin{equation}
\begin{split}
-\frac{dE}{dt}\bigg{|}_{I_{ijklm} \& J_{ijkl}} =& \frac{1168346}{43659} \bigg(\frac{m}{M}\bigg)^{2} v^{16}\bigg(1-\frac{20 S_1 v^3}{3 M^2}-\frac{5 M_2 v^4}{M^3}+\frac{20 S_1^2 v^6}{M^4}+\frac{10 v^7 \left(3 M_2 S_1+M S_3\right)}{M^5}\bigg)\,.
\end{split}
\end{equation}
\end{widetext}
Since we are considering only the first-order contribution of tidal deformability, we limit the expansion of $\Delta N$ to the fifth power of $v$. This is why we need to know the expression of luminosity only up to the twentieth power of $v$, as can be inferred from its relation with $\Delta N$, as given in Eq.~(\ref{N}). But in some cases, as we have calculated the expressions beyond that order, we are showing those expansions here.

In Sec.~\ref{sec:beyond ryan}, these luminosity contributions will be used to find higher order corrections in the expression of phase evolution, beyond what was found by Ryan~\cite{Ryan}.

\section{Luminosity absorbed by the horizon}
\label{sec:Luminosity absorbed inside the horizon}

A complex situation arises when we focus our attention on the GW energy absorbed by the central object when it has a horizon. It is well known that the Teukolsky equation~\cite{Teukolsky,Chandra} can be used to understand the perturbative solutions of the metric. For absorption we study the ingoing solution. The absorbed luminosity can be calculated from there. But the whole problem depends on two things: (a) the central object's vacuum spacetime and (b) the perturbation equation and its solution for that metric. If we knew all possible SAVAR metric solutions, then by solving for the absorption in each one we can identify the effect in those spacetimes. So, in principle, we have to know all such metrics and their contributions. But we can not do that at present because (a) we can not claim that we know all the metric solutions  at the present time and (b) for the solutions that are known these results have not been completely worked out. 

Here we will address the problem in a different manner. The approach we take is heuristic, which needs further and detailed investigation. But as a first step, this is the best we can do. Tagoshi et al.~\cite{absorption} have calculated the luminosity absorbed by the Kerr black hole. We know that all  Geroch-Hansen multipole moments that are non-zero for a general axisymmetric solution are non-zero for Kerr~\cite{Hansen}. We also know that the moments of Kerr depend only on mass and the rotation parameter through $M_l +iS_l = M(ia)^l$~\cite{Hansen}. With this result in hand we can express the luminosity absorbed by a Kerr black hole completely in terms of its multipole moments.

To implement this idea we introduce 
a contribution $-(dE/dt)_H$ to the total luminosity lost from the orbit~\cite{absorption,Tagoshi}:
\begin{widetext}
\begin{equation}
\label{Horizon luminosity}
\begin{split}
-\Bigg(\frac{dE}{dt}\Bigg)_{H} &= \frac{32}{5} \Bigg(\frac{m}{M}\Bigg)^2 v^{15} H \Bigg[-\frac{\chi}{4} - \frac{3\chi^3}{4} -\bigg(\chi + \frac{33}{16}\chi^3\bigg)v^2+ \bigg(2\chi B_2 + \frac{1}{2} + \frac{13}{2}\kappa \chi^2 + \frac{35}{6}\chi^2 - \frac{\chi^4}{4} + \frac{\kappa}{2} + 3\chi^4 \kappa + 6\chi^3 B_2\bigg)v^3\\ 
&+ \bigg(-\frac{43}{7}\chi -\frac{17}{56}\chi^5 - \frac{4651}{336}\chi^3\bigg)v^4 + \bigg(\frac{433}{24}\chi^2 - \frac{95}{24}\chi^4 + 2 - \frac{3}{4}\chi^3B_1+ 2\kappa + \frac{33}{4}\chi^4 \kappa + 6\chi B_2 + 18\chi^3B_2 + \frac{163}{8} \chi^2\kappa\\
&+ \chi B_1\bigg)v^5 + O(v^6)\Bigg],
\end{split}
\end{equation}
\end{widetext}
\begin{align}
B_n = \frac{1}{2i} \Bigg[{\psi}^{(0)} \bigg(3 + \frac{ni\chi}{\sqrt{1 - \chi^2}}\bigg) - {\psi}^{(0)} \bigg(3 - \frac{ni\chi}{\sqrt{1 - \chi^2}}\bigg) \Bigg],
\end{align}
where $H$ is a ``horizon'' parameter, $\chi = \frac{a}{M}$, $\kappa = \sqrt{1-\chi^2}$ and $\psi^{(n)}(z)$ is the polygamma function. The above luminosity term contributes to the total luminosity when the central object has a horizon; in that case $H=1$. When there is no horizon and zero energy absorption by the object, one has  $H=0$ and this term does not contribute. 
In case of partial absorption, which is possible for certain ultracompact objects 
~\cite{mimicing horizon 1,mimicing horizon 2,mimicing horizon 3}, one has $0<H<1$.

$\chi = \frac{a}{M}$ reveals that $\chi^{2s} = \big(\frac{a}{M}\big)^{2s}$. Note that the $a^{2s}$ term can arise only from a very few places. One is from $M_{2s}$ and another is from the multiplication of lower multipole moments. And the same goes for the $\chi^{2s+1}$, which has the main contribution from $S_{2s+1}$. Considering all such aspects we can write $\chi^m$ in terms of the multipole moments as follows:
\begin{align}
\label{S1} \chi &= \frac{S_1}{M^2},\\
\chi^2 &= -a_2 \frac{M_2}{M^3} + a_1 \frac{S_1^2}{M^4},\\
\chi^3 &= a_3 \bigg(\frac{S_1}{M^2}\bigg)^3 +a_4\frac{S_1}{M^2}\bigg\{-a_2 \frac{M_2}{M^3} + a_1 \frac{S_1^2}{M^4}\bigg\}  - a_5 \frac{S_3}{M^4},
\end{align}
\begin{widetext}
\begin{equation}
\begin{split}
\chi^4 &= a_6 \bigg(\frac{S_1}{M^2}\bigg)^4 +a_7\bigg(\frac{S_1}{M^2}\bigg)^2\bigg\{-a_2 \frac{M_2}{M^3} + a_1 \frac{S_1^2}{M^4}\bigg\}+a_8 \frac{S_1}{M^2}\bigg[a_3 \bigg(\frac{S_1}{M^2}\bigg)^3 +a_4\frac{S_1}{M^2}\bigg\{-a_2 \frac{M_2}{M^3} + a_1 \frac{S_1^2}{M^4}\bigg\}  - a_5 \frac{S_3}{M^4}\bigg]\\ &+a_9\bigg[-a_2 \frac{M_2}{M^3} + a_1 \frac{S_1^2}{M^4}\bigg]^2+ a_{10} \frac{M_4}{M^5},
\end{split}
\end{equation}
\begin{equation}
\begin{split}
\chi^5 &= \frac{S_1}{M^2}\bigg[a_{11}\bigg(\frac{S_1}{M^2}\bigg)^4 +\bigg\{-a_2 \frac{M_2}{M^3} + a_1 \frac{S_1^2}{M^4}\bigg\}\bigg\{a_{12} \bigg(\frac{S_1}{M^2}\bigg)^2+ a_{13}\bigg(-a_2 \frac{M_2}{M^3} + a_1 \frac{S_1^2}{M^4}\bigg)\bigg\} +a_{15}\bigg\{a_6 \bigg(\frac{S_1}{M^2}\bigg)^4\\
&+a_7\bigg(\frac{S_1}{M^2}\bigg)^2\bigg\{-a_2 \frac{M_2}{M^3}+ a_1 \frac{S_1^2}{M^4}\bigg\}+a_8 \frac{S_1}{M^2}\bigg[a_3 \bigg(\frac{S_1}{M^2}\bigg)^3 +a_4\frac{S_1}{M^2}\bigg\{-a_2 \frac{M_2}{M^3}+ a_1 \frac{S_1^2}{M^4}\bigg\}  - a_5 \frac{S_3}{M^4}\bigg]+a_9\bigg\{-a_2 \frac{M_2}{M^3} + a_1 \frac{S_1^2}{M^4}\bigg\}^2\\
&+ a_{10} \frac{M_4}{M^5}\bigg\}\bigg]+\bigg[a_3 (\frac{S_1}{M^2})^3 +a_4\frac{S_1}{M^2}\{-a_2 \frac{M_2}{M^3}+ a_1 \frac{S_1^2}{M^4}\}  - a_5 \frac{S_3}{M^4}\bigg]\bigg[a_{14}(\frac{S_1}{M^2})^2 +a_{16}\{-a_2 \frac{M_2}{M^3} + a_1 \frac{S_1^2}{M^4}\}\bigg]+a_{17}\frac{S_5}{M^6},
\end{split}
\end{equation}
\end{widetext}
where the $a_i$s are 17 undetermined parameters, and are to be distinguished from $a$, which is the spin parameter and, contrastingly, does not have an index. The aforementioned equations have been formed by finding in how many ways $\chi^l$ can be constructed from the $M_L$ and $S_L$. While doing that we only focused on how they can be constructed by multiplying different moments. After that those contributions have been added with the introduction of the $a_i$s. However, as both the RHS and the LHS of the corresponding equations should be equal to $\chi^l$, these $a_i$s are not all independent. They satisfy four consistency equations, so there are 13 undetermined parameters. The equations satisfied by them are,
\begin{align}
a_1 + a_2 = &1,\\
a_3 + a_4 + a_5 = &1,\\
a_6 + a_7 + a_8 + a_9 + a_{10} = &1,\\
a_{11} + a_{12} + a_{13} + a_{14} + a_{15} + a_{16}+a_{17} = &1.
\end{align}
A measurement of the absorbed luminosity for an SAVAR metric that is not Kerr can help constrain these parameters for that space-time especially, if the moments from the precession frequencies can also be found.

The expression is very complicated, but if we only consider the dominant contribution from each moment, as Ryan did, then it becomes much simpler. We now, however, opt to be as rigorous as possible.

For future purpose reexpress Eq.(\ref{Horizon luminosity}) in terms of five new parameters:
\begin{equation}
\begin{split}
-\Bigg(\frac{dE}{dt}\Bigg)_H &= \frac{32}{5} \Bigg(\frac{m}{M}\Bigg)^2 v^{15} H \Bigg[A^{\prime} + B^{\prime} v^2 + C^{\prime} v^3 + D^{\prime} v^4\\
&+ E^{\prime} v^5\Bigg].
\end{split}
\end{equation}
where $A^{\prime}$,...,$E^{\prime}$ are newly defined expansion coefficients. Though the above expression depends on terms that are multiples of the different multipole moments, it is understandable that in case of Kerr it becomes much simpler. If the only involved parameters related to the central object are mass and angular momentum, then due to the uniqueness theorem the external metric will be Kerr if it is a black hole. In that case the absorption will depend only on $a$ and $M$. But for the Kerr family, dependencies of the moments on mass and angular momentum are very simple. Owing to that these multipole moments are directly related to each other, and it does not matter if we side-step the ambiguity of the values of different $a_i$s. Therefore, we can choose:
\begin{equation}
a_2 = a_5 = a_{10} = a_{17} = 1.
\end{equation}
And all other $a_i = 0$. For the Kerr metric this will not change the result at all. Therefore, when there are no free parameters other than $M$ and $a$, one can take this simple form. 
So, the luminosity absorbed by the horizon becomes, 
\begin{widetext}
\begin{equation}
\begin{split}
-\Bigg(\frac{dE}{dt}\Bigg)_H &= \frac{32}{5} \Bigg(\frac{m}{M}\Bigg)^2 v^{15} H \Bigg[-\frac{S_1}{4 M^2} + \frac{3 S_3}{4 M^4} -\bigg(\frac{S_1}{M^2} - \frac{33 S_3}{16 M^4}\bigg)v^2+ \bigg(2\frac{S_1}{M^2}B_2 + \frac{1}{2} - \frac{13 M_2}{2 M^3}\kappa  - \frac{35 M_2}{6 M^3} -\frac{M_4}{4M^5} + \frac{\kappa}{2}\\
&+3\frac{M_4}{M^5} \kappa - 6\frac{S_3}{M^4} B_2\bigg)v^3+ \bigg(-\frac{43 S_1}{7 M^2} -\frac{17S_5}{56 M^6} + \frac{4651 S_3}{336M^4}\bigg)v^4 + \bigg(-\frac{433M_2}{24M^3} - \frac{95M_4}{24M^5} + 2 + \frac{3 S_3}{4 M^4}B_1\\ 
&+ 2\kappa + \frac{33 M_4}{4M^5} \kappa + 6\frac{S_1}{M^2}B_2 - 18\frac{S_3}{M^4}B_2 - \frac{163M_2}{8M^3}\kappa + \frac{S_1}{M^2}B_1 \bigg)v^5 + O(v^6)\Bigg],\\
B_n &= \frac{1}{2i} \Bigg[{\psi}^{(0)} \bigg(3 + \frac{ni S_1}{\sqrt{M^4 + M_2 M}}\bigg)- {\psi}^{(0)} \bigg(3 - \frac{ni S_1}{\sqrt{M^4 + M_2 M}}\bigg) \Bigg],
\end{split}
\end{equation}
\end{widetext}
where, $\kappa = \sqrt{1+\frac{M_2}{M^3}}$. This simplification happens because the relation between various multipole moments and the powers of $\chi$ become simple for the Kerr metric. Their expressions are,
\begin{align}
\chi &= \frac{S_1}{M^2},\\
\chi^2 &=  -\frac{M_2}{M^3},\\
\chi^3 &= - \frac{S_3}{M^4},\\
\chi^4 &= \frac{M_4}{M^5},\\
\chi^5 &= \frac{S_5}{M^6} .
\end{align}
The basic idea employed here is to use horizon absorption as evidence that the central object has a horizon. Recently, Maselli et al.~\cite{H} have used this idea for the same purpose. They too introduced an absorption coefficient~($\gamma$), which is identical to our $H$. The significance of this term was arrived at independently~\cite{IAGRG}.

\section{Complete mass-only dependent part through $v^{10}$ }
\label{sec:mass dep. part}

Ryan had included mass-dependent terms in the luminosity that resulted from the perturbation of the Schwarzschild black hole. As the only intention of that work was to look into the observational aspects of the No-hair theorem, it was good enough to consider them up to fourth power of $v$. But the main purpose of the present work is to include tidal effect and absorption by the central object. Since the tidal contribution occurs at much higher order of $v$ we need to know the luminosity up to that power of $v$ beyond lowest order. For that reason we are including this correction up to tenth power of $v$~\cite{Tagoshi}:
\begin{widetext}
\begin{equation}
\begin{split}
-\Bigg(\frac{dE}{dt}\Bigg)_M =& \frac{32}{5}{\bigg(\frac{m}{M}\bigg)}^2 v^{10}\bigg[1-\frac{1247}{336}v^2 + 4\pi v^3 - \frac{44711}{9072}v^4-\frac{1712}{105} \ln v\,\, v^6+ \frac{232597}{4410} \ln v\,\, v^8- \frac{6848}{105}\pi \ln v\,\, v^9\\
&+ \frac{916628467}{7858620} \ln v\,\, v^{10}+ \alpha v^5 + \beta v^6 + \nu v^7 + \delta v^8 + \epsilon v^9 + \phi v^{10}\bigg]\,.
\end{split}
\end{equation}
\end{widetext}
For future purpose we have expressed the expansion with some newly introduced parameters. The original expression is given below; comparing it with the above we can easily find those parameters. The full expression for the luminosity is~\cite{Tagoshi},
\begin{widetext}
\begin{equation}
\begin{split}
&-\Bigg(\frac{dE}{dt}\Bigg)_M\\
=& \frac{32}{5}{\bigg(\frac{m}{M}\bigg)}^2 v^{10}\bigg[1-\frac{1247}{336}v^2 + 4\pi v^3 - \frac{44711}{9072}v^4 - \underbrace{\frac{8191}{672} {\pi}}_{= -\alpha} v^5
+ \bigg(\underbrace{\frac{6643739519}{69854400} - \frac{1712}{105}\gamma + \frac{16}{3} \pi^2 - \frac{3424}{105}\ln 2}_{= \beta} -\frac{1712}{105} \ln v\bigg)v^6\\
&- \underbrace{\frac{16285}{504}\pi}_{= -\nu} v^7
+ \bigg(\underbrace{-\frac{323105549467}{3178375200} + \frac{232597}{4410} \gamma - \frac{1369}{126}\pi^2 + \frac{39931}{294} \ln 2 - \frac{47385}{1568} \ln 3}_{= \delta} + \frac{232597}{4410} \ln v\bigg)v^8\\
&+ \bigg(\underbrace{\frac{265978667519}{745113600}\pi - \frac{6848}{105}\pi \gamma - \frac{13696}{105}\pi \ln 2}_{=\epsilon} - \frac{6848}{105}\pi \ln v\bigg)v^9\\
+& \bigg(\underbrace{-\frac{2500861660823683}{2831932303200} + \frac{916628467}{7858620}\gamma - \frac{424223}{6804}\pi^2 -\frac{83217611}{1122660}\ln 2 + \frac{47385}{196}\ln 3}_{= \phi} + \frac{916628467}{7858620} \ln v \bigg)v^{10}\bigg]\,,
\end{split}
\end{equation} 
\end{widetext}
where $\gamma$ is the Euler constant.
Since this expression is independent of all the other multipole moments, apart from the mass, this contribution will be present in all  SAVAR metrics. 

\section{Motion of the central object }
\label{sec:central motion}

Another contribution that becomes important in this calculation is the effect due to the motion of the central object. This was mentioned by Ryan~\cite{Ryan}, and the results necessary for his calculation were presented there. We will take those basic results and identify the terms that will be important for our purpose.\\

Let the axis of symmetry of the SAVAR metric be denoted by the vector $\tilde{z}$, which can be defined in terms of the Killing vector corresponding to this symmetry~\cite{Hansen}. If $\Lambda$ represents the spatial infinity then we have~\cite{Hansen},
\begin{align}
{\tilde{z}}_b{\tilde{z}}^b|_{\Lambda} = 1\,.
\end{align}
Since the metric is  asymptotically flat, the axial Killing vector generates rotation on tensors at $\Lambda$. The moments should be rotationally invariant. But the only tensors at $\Lambda$ that are invariant under the action of the axial Killing vector are the ones that are outer products of the metric and $\tilde{z}$; so, the $2^s$ moments have to be multiples of the symmetric, trace-free (STF) outer product of $\tilde{z}$ with itself \big($\big[{\tilde{z}}_{b_1}...{\tilde{z}}_{b_s}\big]^{\rm STF}{\big|}_{\Lambda}$\big). 

The definition of a STF tensor can be generalized following Thorne's expression~\cite{Thorne},
\begin{align}
A^{\rm sym}_{b_1 ...b_s} = [A_{b_1 ...b_s}]^S = \frac{1}{l!}\sum_{\pi}A_{b_{\pi (1)} ...b_{\pi (s)}}\,,
\end{align}
where $A^{\rm sym}_{b_1 ...b_s}$ is the completely symmetrized part of $A_{b_1 ...b_s}$ and $\pi$ represents all possible permutations of its indices. Now the symmetric, trace-free part can easily be found from Thorne's expression~\cite{Thorne},
\begin{equation}
\begin{split}
[A_{b_1 ...b_s}]^{\rm STF} =& \sum_{n=0}^{{\rm Floor}(\frac{s}{2})}\frac{(-1)^n s! (-2 n+2 s-1)\text{!!}}{(2 n)\text{!!} (2 s-1)\text{!!} (s-2 n)!}\\
&\times \delta_{(b_1 b_2} ...\delta_{b_{2n-1}b_{2n}}A^{\rm sym}_{b_{2n+1} ...b_s)j_1j_1 ...j_n j_n}\,,
\end{split}
\end{equation}
where the repeated indices $j_k$ are  contracted over and index symmetrization is defined as, $B_{(i}C_{j)} \equiv \frac{1}{2}( B_i C_j + B_j C_i)$.
The definitions of the $2^s$ moments ($M_L$ and $S_L$) can be found in \cite{Hansen}. Since only the axis vector and the metric remain invariant under rotation, the $2^s$ moments are determined by the numbers $M_s$ and $S_s$ defined as~\cite{Hansen}:
\begin{align}
\label{M2}M_{s} =& \frac{1}{s!}M_{{b_1}...b_s}{\tilde{z}}^{{b_1}}...{\tilde{z}}^{b_s}|_{\Lambda},\\
\label{S2}S_{s} =& \frac{1}{s!}S_{{b_1}...b_s}{\tilde{z}}^{{b_1}}...{\tilde{z}}^{b_s}|_{\Lambda}\,,
\end{align}
where $s$ belongs to the set of positive integer numbers. But for this work we need to know the $2^s$ moments in terms of $M_s$ and $S_s$. Since the moments will be combinations of the outer products of the axis vector, the $2^s$ moments will be,
 \begin{align}
 \label{mass moment}M_{b_1 ...b_s} =& \alpha_{M_s} \big[{\tilde{z}}_{b_1}...{\tilde{z}}_{b_s}\big]^{\rm STF}{\big|}_{\Lambda},\\
 \label{current moment}S_{b_1 ...b_s} =& \alpha_{S_s} \big[{\tilde{z}}_{b_1}...{\tilde{z}}_{b_s}\big]^{\rm STF}{\big|}_{\Lambda},\,
 \end{align}
where $\alpha_{M_s}$ and $\alpha_{S_s}$ are some numbers yet to be determined. We can put Eqs.~(\ref{mass moment}) and (\ref{current moment}) into Eqs.~(\ref{M2}) and (\ref{S2}) in order to find these numbers in terms of $M_s$ and $S_s$:
\begin{align}
M_{s} =& \frac{\alpha_{M_s}}{s!}T_s,\\
S_{s} =& \frac{\alpha_{S_s}}{s!}T_s\,,
\end{align}
where
\begin{equation}
\begin{split}
T_s &\equiv [\tilde{z}_{b_1}...\tilde{z}_{b_s}]^{\rm STF}\tilde{z}^{b_1}...\tilde{z}^{b_s}\\
&= \sum_{n=0}^{{\rm Floor}(\frac{s}{2})}\frac{(-1)^n s! (-2 n+2 s-1)\text{!!}}{(2 n)\text{!!} (2 s-1)\text{!!} (s-2 n)!}\,.
\end{split}
\end{equation}
Hence, 
\begin{align}
S_1 =& \alpha_{S_1}\\
S_{b_1} =& S_1 {\tilde{z}}_{b_1}\\
M_{2} =& \frac{\alpha_{M_2}}{2!} \frac{2}{3}\,,\\
M_{b_1 b_2} =& 3M_2 \big[{\tilde{z}}_{b_1}{\tilde{z}}_{b_2}\big]^{\rm STF}{\big|}_{\Lambda}\,.
\end{align}
If the orbiting companion were absent, then the moment of the system would have been determined by the stationary moment of the central body alone. So, there would have been no radiation. In reality, due to the orbiting companion the larger object will move along a path $\sim - \big(m / M \big) x_k$ in the center of mass frame, where $x_k$ is the smaller companion's position. Therefore, the multipolar contribution due to the ``moving" large mass would be the stationary moment displaced by $\big(m/ M \big) x_k$. Ryan already had included the contribution of $S_1$ due to this effect~\cite{Ryan}. 

It is now simple to see that the only other contribution through the tenth power of $v$ will arise from $M_2$. This is because of the number of time-derivatives on the radiative moment and the number of position vectors present in each term. Since we are assuming a circular orbit, we can write the smaller companion's position as,
\begin{align}
\label{position}x_1 = \rho \cos(\Omega t),\,\,\,x_2 = \rho \sin(\Omega t),\,\,\,x_3 = 0\,,
\end{align}
where $\rho$ is the separation between the two bodies.
Because of the motion of the central object the radiative moments get corrected by~\cite{Ryan},
\begin{align}
\delta I_{L+1} &= [-(l+1) I_L \big(m/ M \big) x_{b_{l+1}}]^{\rm STF},\\
\delta J_{L+1} &= [-\frac{l(l+2)}{l+1}J_L \big(m/ M \big) x_{b_{l+1}}]^{\rm STF}.
\end{align}
Therefore, we find
\begin{align}
\label{Iijk}I_{ijk} =\big[ m x_i x_j x_k - 9M_2 \big[{\tilde{z}}_i{\tilde{z}}_j\big]^{\rm STF} \frac{m}{M} x_k\big]^{\rm STF},
\end{align}
\begin{align}
\label{Jij}J_{ij} = \big[m x_i \epsilon_{jkl} x_k \frac{dx_l}{dt} - \frac{3m}{2M} x_j S_1 {\tilde{z}}_i\delta_{j3}\big]^{\rm STF}.
\end{align}
Substituting Eq.~(\ref{position}) into Eq.~(\ref{Iijk}) and inserting the resulting expression in Eq.~(\ref{time avg}), followed by a separation of the contribution of the small mass calculated above, we find the extra contribution to be:
\begin{equation}
\begin{split}
-\frac{dE}{dt}\bigg{|}_{I_{ijk},{\rm C.M.}} =& \frac{32}{5}\bigg(\frac{m}{M}\bigg)^{2} v^{10}\bigg[\frac{M_2 v^6}{336 M^3}-\frac{v^9 \left(M_2 S_1\right)}{126 M^5}\\
&+\frac{M_2^2 v^{10}}{84 M^6}+O\left(v^{11}\right)\bigg]\,,
\end{split}
\end{equation}
where C.M. is the short-hand for central body's motion. Similarly, putting Eq.~(\ref{position}) into Eq.~(\ref{Jij}) and using the result in Eq.~(\ref{time avg}) and separating the small mass' contribution calculated earlier we deduce the extra contribution of $J_{ij}$, complete through tenth power beyond the lowest order, as
\begin{widetext}
\begin{equation}
\label{luminosity due to J_ij C.M.}
\begin{split}
-\frac{dE}{dt}\bigg{|}_{J_{ij}, \rm C.M.} =& \frac{32}{5}\bigg(\frac{m}{M}\bigg)^{2} v^{10}\bigg[-\frac{S_1 v^3}{12 M^2}+\frac{S_1^2 v^4}{16 M^4}+\frac{2 S_1^2 v^6}{9 M^4}-\frac{v^7 \left(S_1^3-2 M M_2 S_1\right)}{12 M^6} -\frac{v^8 \left(M_2 S_1^2\right)}{16 M^7}\\
&-\frac{2 S_1^3 v^9}{9M^6}+\frac{S_1 v^{10} \left(-12 M^2 S_3-12 M M_2 S_1+S_1^3\right)}{36 M^8}\bigg].
\end{split}
\end{equation}
\end{widetext}
The first two terms inside the square brackets of the RHS were calculated in Refs.~\cite{kidder} and \cite{Ryan}.

With Eq.(\ref{luminosity due to J_ij C.M.}), we have now finished calculating all the terms arising from the relevant effects, up to the order we need. This sets the stage for calculating how much the phasing will get modified owing to the aforementioned contributions.

\section{Corrections beyond Ryan and their measurability}
\label{sec:beyond ryan}
\subsection{Corrections}
In previous sections we gave the expressions of all possible contributions to the luminosity of an EMRI, namely:
\begin{widetext}
\begin{equation}
\begin{split}
-\frac{dE}{dt}\bigg{|}_{\rm Total} =& -\frac{dE}{dt}\bigg{|}_{I_{ijk}, J_{ij}} - \frac{dE}{dt}\bigg{|}_{I_{ij}} - \frac{dE}{dt}\bigg{|}_M - \frac{dE}{dt}\bigg{|}_H- \frac{dE}{dt}\bigg{|}_{\rm Tidal}- \frac{dE}{dt}\bigg{|}_{I_{ijkl},J_{ijk}}- \frac{dE}{dt}\bigg{|}_{I_{ijklm},J_{ijkl}}\\
&- \frac{dE}{dt}\bigg{|}_{I_{ijk},{\rm C.M.}}- \frac{dE}{dt}\bigg{|}_{J_{ij},{\rm C.M.}}\,.
\end{split}
\end{equation}
\end{widetext}
The expression for $\Delta E$ needed for our calculation was obtained nearly completely by Ryan~\cite{Ryan}. The net result below is the combination of Ryan's result and the contribution due to the tidal interaction, which we calculated in Eq.~(\ref{Tidal Delta E}). Thus,
\begin{widetext}
\begin{equation}
\begin{split}
\label{delta E}\frac{\Delta E }{m}=&\frac{v^2}{3}-\frac{v^4}{2} +\frac{20 S_1 v^5}{9 M^2}+ \left(\frac{M_2}{M^3}-\frac{27}{8}\right) v^6+\frac{28 S_1 v^7}{3 M^2}+v^8 \left(\frac{80 S_1^2}{27 M^4}+\frac{70 M_2}{9 M^3}-\frac{225}{16}\right)+v^9 \left(\frac{6M_2 S_1}{M^5}-\frac{6 S_3}{M^4}+\frac{81 S_1}{2 M^2}\right)\\
&+v^{10} \left(\frac{35 M_2^2}{12 M^6}-\frac{35 M_4}{12 M^5}+\frac{115 S_1^2}{18 M^4}+\frac{935 M_2}{24M^3}-\frac{6615}{128}\right)+v^{11} \left(\frac{1408 S_1^3}{243 M^6}+\frac{968 M_2 S_1}{27 M^5}-\frac{352 S_3}{9 M^4}+\frac{165 S_1}{M^2}\right)\\
&+v^{12} \left(\frac{24 M_2S_1^2}{M^7}-\frac{24 S_1 S_3}{M^6}+\frac{93 M_2^2}{4 M^6}-\frac{99 M_4}{4 M^5}-\frac{123 S_1^2}{14 M^4}+\frac{9147 M_2}{56 M^3}-\frac{45927}{256}+\frac{X}{m}\right)+...,\\
\end{split}
\end{equation}
\end{widetext}
which includes all the terms we set out to find.

To make the expression of $\Delta N$ simple, it helps to define the following variables:
\begin{equation}
\begin{split}
A_1 =& \alpha + A^{\prime}H,\\
B_1 =& \nu + B^{\prime}H,\\
C_1 =& \delta + C^{\prime}H,\\
D_1 =& \epsilon + D^{\prime}H,\\
E_1 =& \phi + E^{\prime}H,
\end{split}
\end{equation}
as well as,
\begin{align}
\begin{split}
A\,\, =& \frac{M^2}{m^2}\frac{6{\eta}^2}{M_T^5}\big\{\frac{m + 3M}{m}{\lambda}_{m}+\frac{M + 3m}{M}{\lambda}_{M}\big\}, \\
X\,\, =& -18\frac{\eta }{M^4{_T}} \big\{{\frac{M}{m}}{\lambda}_{m}+{\frac{m}{M}}{\lambda}_{M}\big\},
\end{split}
\end{align}
where $A^{\prime},B^{\prime}, C^{\prime}, D^{\prime}$, and $E^{\prime}$ have been defined in Sec.~\ref{sec:Luminosity absorbed inside the horizon} and $A$ and $X$ have been defined in Sec.~{\ref{sec:tides}}.
The Greek parameters~(except $\lambda$ and parameters depending on it) depend only on mass. Terms containing $H$ are present when the central object has a horizon.

Now we have the full expression for $\Delta E$ and the total luminosity. Therefore, from these quantities we can find the phase evolution. Putting Eq.~(\ref{delta E}) and the total rate of energy contribution in Eq.~(\ref{N}), the result we obtain is as follows:

\begin{equation}\label{N series}
\Delta N = \frac{5 }{96 \pi  q v^5}\sum_{n =0}^{10}N_n v^n,
\end{equation}
where $q = m/M$.
The expressions for $N_n$ can be found in \ref{expressions}.

With the deviation of the expression of the phase evolution we have achieved the goal we had set out for. Usefulness of this result is manyfold, as will be discussed in the later sections.

 The phase can be calculated as \cite{phase expression},
\begin{equation}\label{phase}
\begin{split}
\psi(f) = 2\pi f t_c - 2\phi_c-\frac{\pi}{4} +  I(v) -  I(v_i)
\end{split}
\end{equation}

\begin{equation}
I(v) - I(v_i) = \int^v_{v_i} d\bar{v} (v^3-\bar{v}^3) \frac{6\pi\Delta N}{\bar{v}^4}
\end{equation}

\begin{equation}\label{I series}
I(v) = \sum_{n=-5}^{5}I_n (v) v^n,
\end{equation}
where the form of $I_n(v)$ can be found in \ref{expressions}.

\subsection{Finding the Horizon }

We now utilize the above results to analyze the challenges involved in deducing from future observations of GWs emitted by EMRIs whether their central object has a horizon or not. 
 
Among all the parameters in the expression of $\Delta N$, the ones denoted by Greek letters~(except $\lambda$) arise from the mass-dependent terms alone, as defined earlier. Terms containing $H$ arise due to absorption by the central object if it has a horizon. Other multipole-dependent terms arise from the higher order corrections and the motion of the central object. From the expressions of $A_1,B_1,C_1,D_1$, and $E_1$, it is noticeable that these parameters depend on the horizon parameter $H$. Therefore, the expression of the phase evolution, in general, depends on a set of multipole moments of the central body, the Horizon parameter, mass of the small compact object and the tidal deformabilities of the two bodies.

Effect of self-force is of order $\epsilon \equiv (\frac{m}{M})$. Therefore, in case  of EMRIs this effect is small. But most of the orbits in EMRI survive long enough without merging; building a cumulative effect of self-force. To have a significant amount of accumulation, the particle should stay in orbit long enough before merging, i.e. of the order of $\frac{1}{\epsilon}$ or longer~\cite{self-force}. But if we consider an event where a compact object comes in from a very large distance, stays on an almost circular orbit for a very short period of time and goes back out to a large distance~\cite{hyperbolic zoom whirl}, the accumulation of self-force correction terms would be small. In these scenarios, one can ignore the contribution of self-force while calculating the expression of $\Delta N$, and use the results derived here.

To determine the values of these variables from observations, it will be essential to use the precession frequencies mentioned earlier~(see Eqs.~(\ref{frequency 1}) and (\ref{frequency 2})) \cite{Li}. In principle, by using the precession frequencies it is possible to deduce the values of the multipole moments of the central body and the mass of the small companion. These values can be used to separate out the influence of these moments on the phase evolution from that of the horizon parameter and the tidal deformabilities of the two bodies. 
 
But the situation can be somewhat simpler. To wit, for a black hole the tidal Love number is zero but $H=1$; therefore, only two of the three aforementioned influences will be present when the central object is a black hole. (Note that we are ignoring the effect of a possible horizon that the smaller object may have on the waveform.) For a general EMRI, whenever $H$ is nonzero for the central object, its tidal Love number will be zero, and {\it vice versa}. 
If both components are black holes, then the situation will become even simpler as there will be no degeneracy left. In this case $H$ is non-zero but tidal deformabilities are zero. 

Few other applications can be thought of. It has been suggested that the behaviour of the tidal deformability in the black hole limit could be used to probe Planckian correction near horizon \cite{H, Andrea 2019}. In certain alternate theories of gravities there are objects that contain a horizon around them but rather than having a zero tidal deformability it takes negative values \cite{Kabir 2019}.

In general, confirming the presence or absence of a horizon will not be simple, owing to possibly competing influences of the aforementioned terms on the waveform phase.
When the central object is not a black hole the degeneracy is between the two deformabilities. If there is a black hole then there is the degeneracy between $H$ and the tidal deformability of the small body. But in GW observations {\it a priori} we will not know what type of components make up the system. Potentially, the signal from one type of system may mimic that from another. For this reason a detailed analysis is needed, which is beyond the scope of this paper and will be reserved for a future work.

\section{Parameter estimation}
\label{sec:paramestim}

To obtain some quantitative sense of how accurately some of the crucial central object parameters will be measurable, we performed a set of Fisher information matrix~\cite{Helstrom,Gair 2017} studies. As we show below, these estimates provide cautious optimism for the possibility that certain tests of the No-hair theorem can be performed with EMRIs in LISA. A more conclusive statement in this regard will have to wait for more accurate modeling of EMRI waveforms, as we have already clarified above.

\begin{widetext}
\begin{figure*}
\includegraphics[width=7.cm]{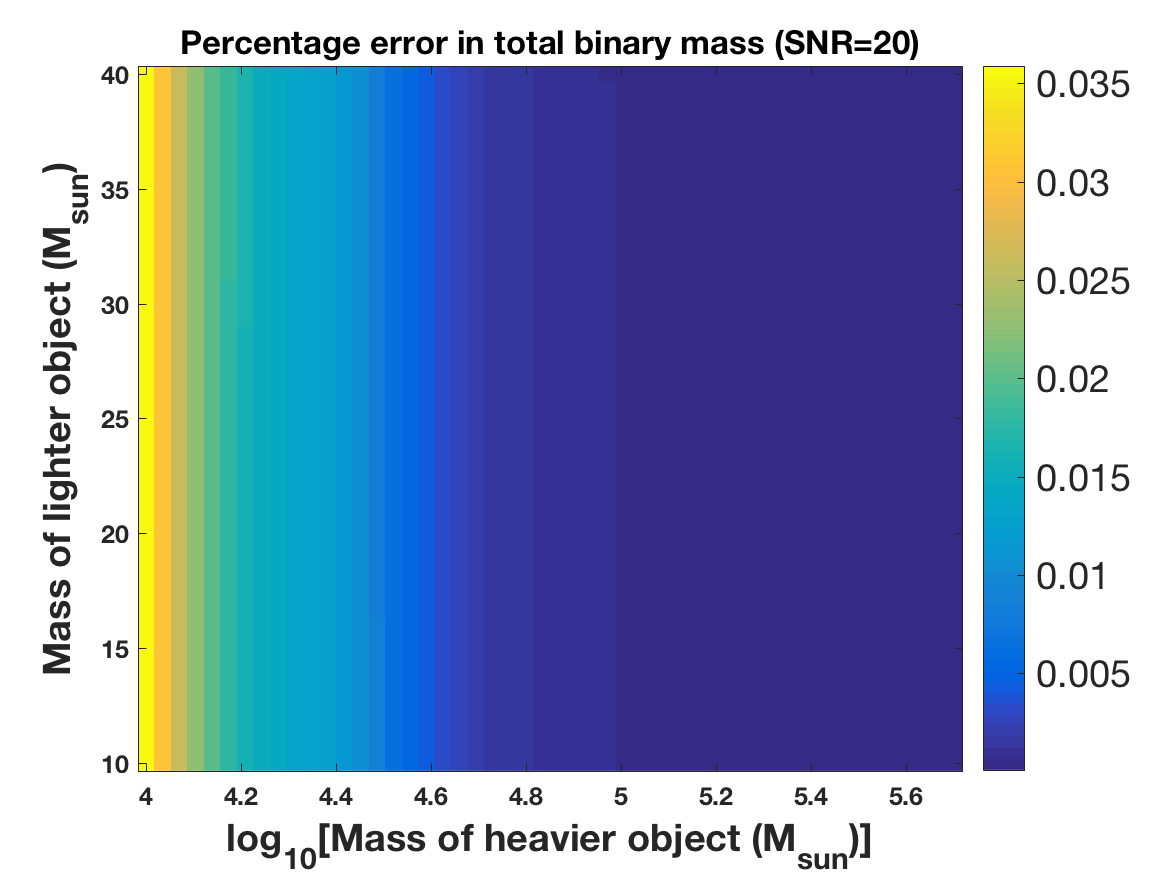}
\includegraphics[width=7.cm]{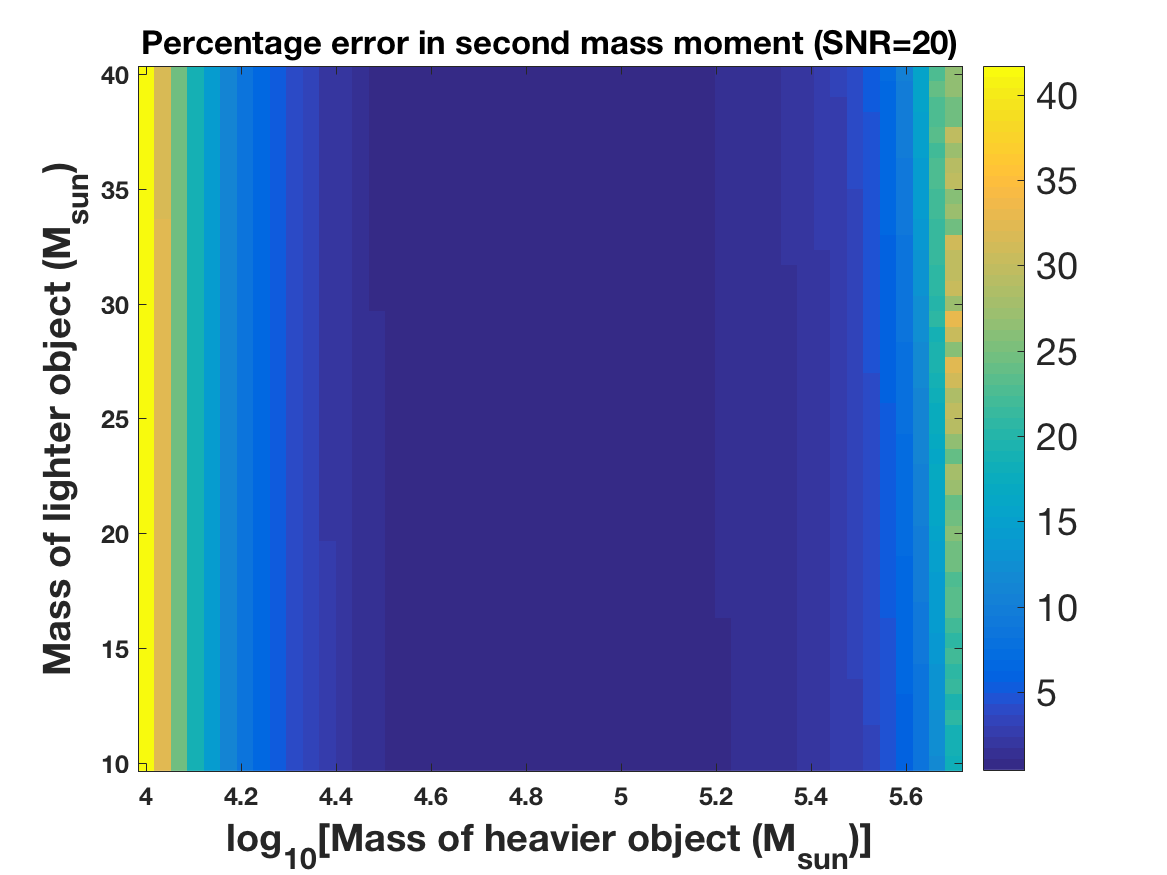}
\includegraphics[width=7.cm]{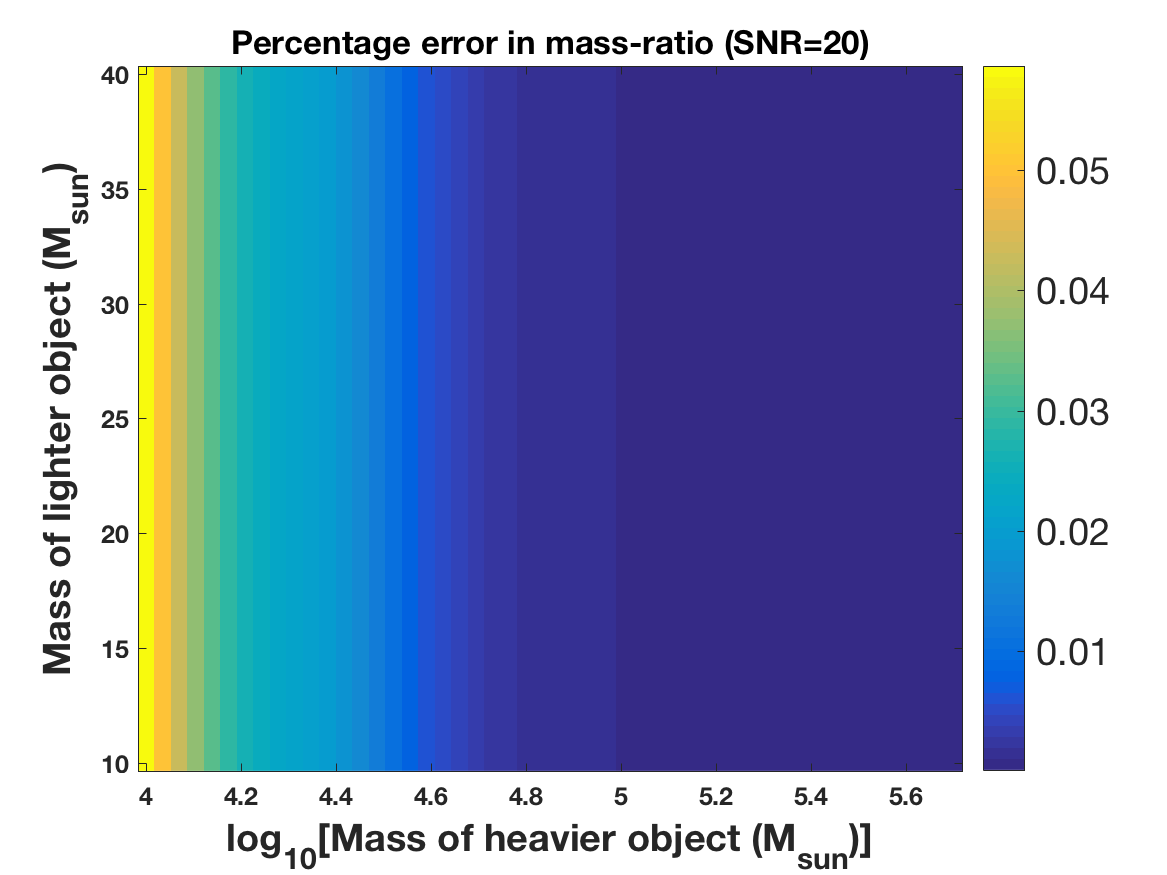}
\includegraphics[width=7.cm]{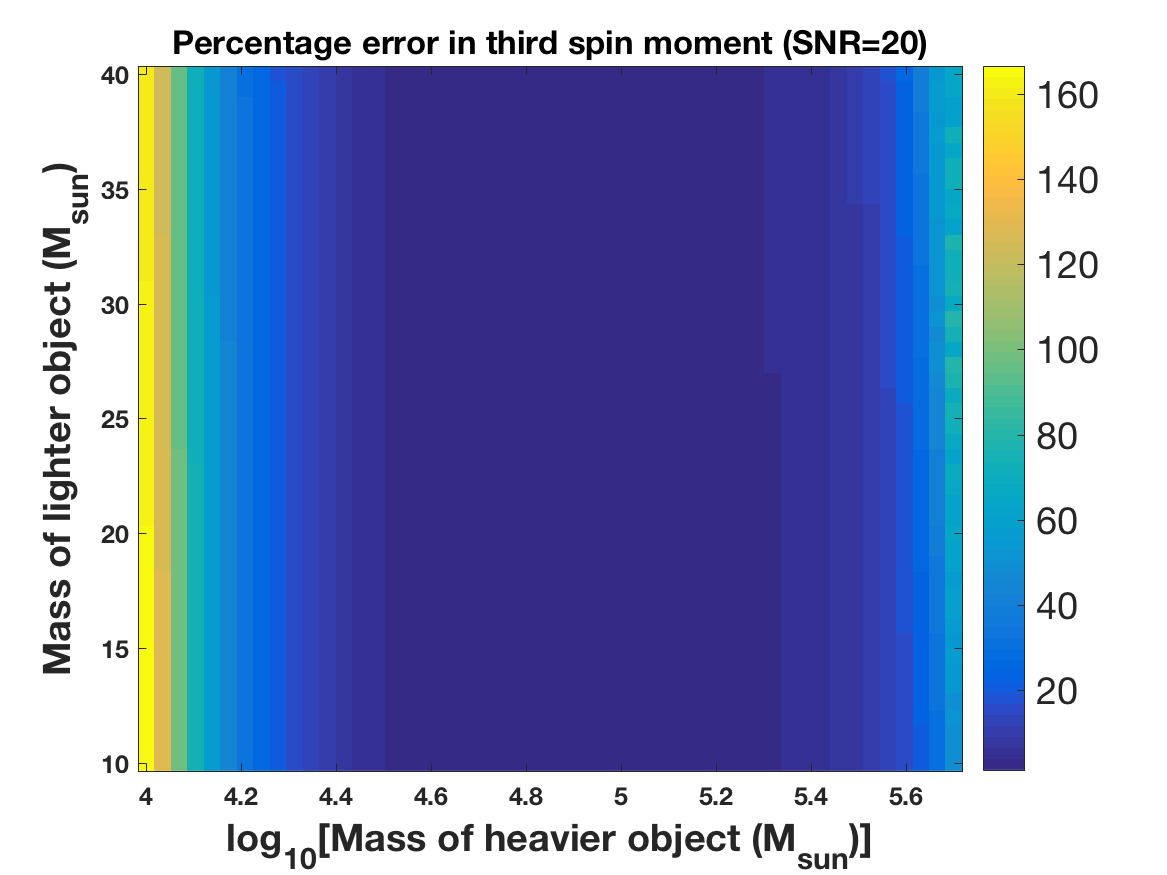}
\includegraphics[width=7.cm]{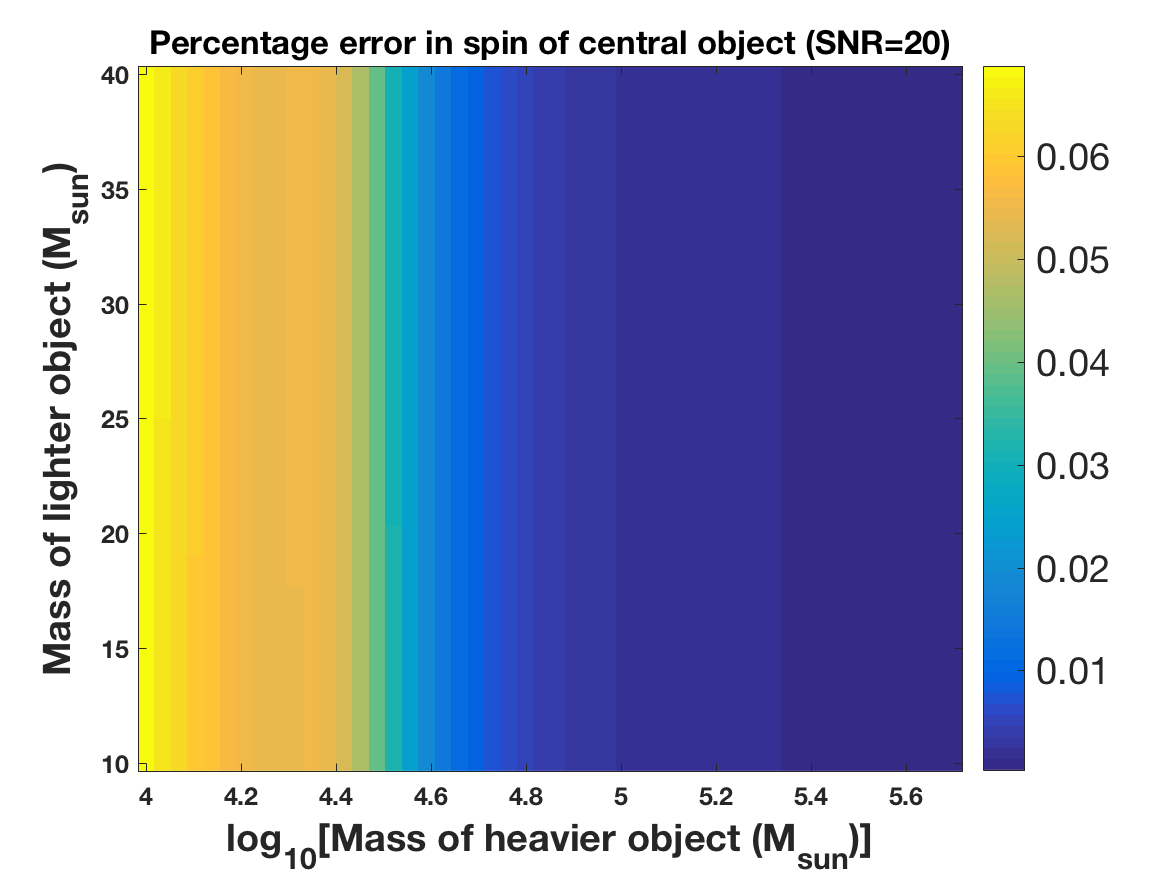}
\includegraphics[width=7.cm]{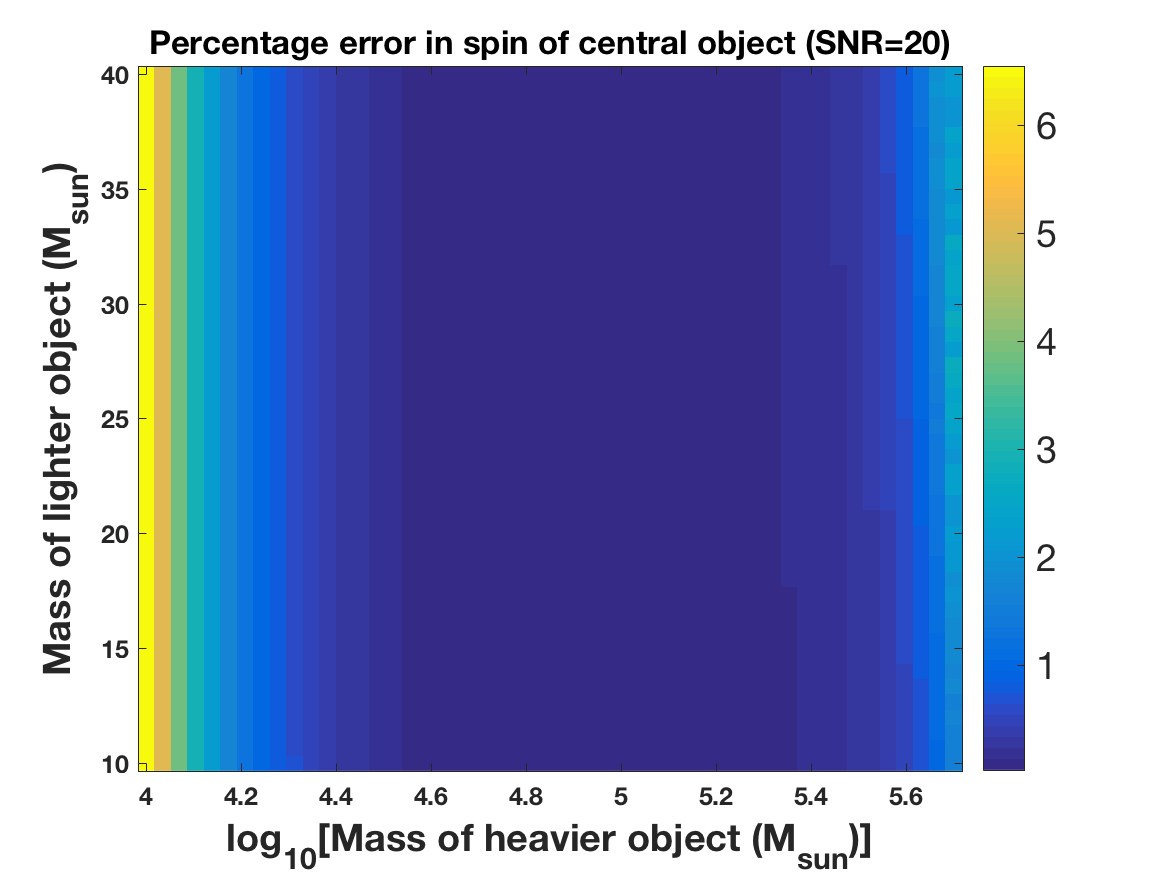}
\includegraphics[width=7.cm]{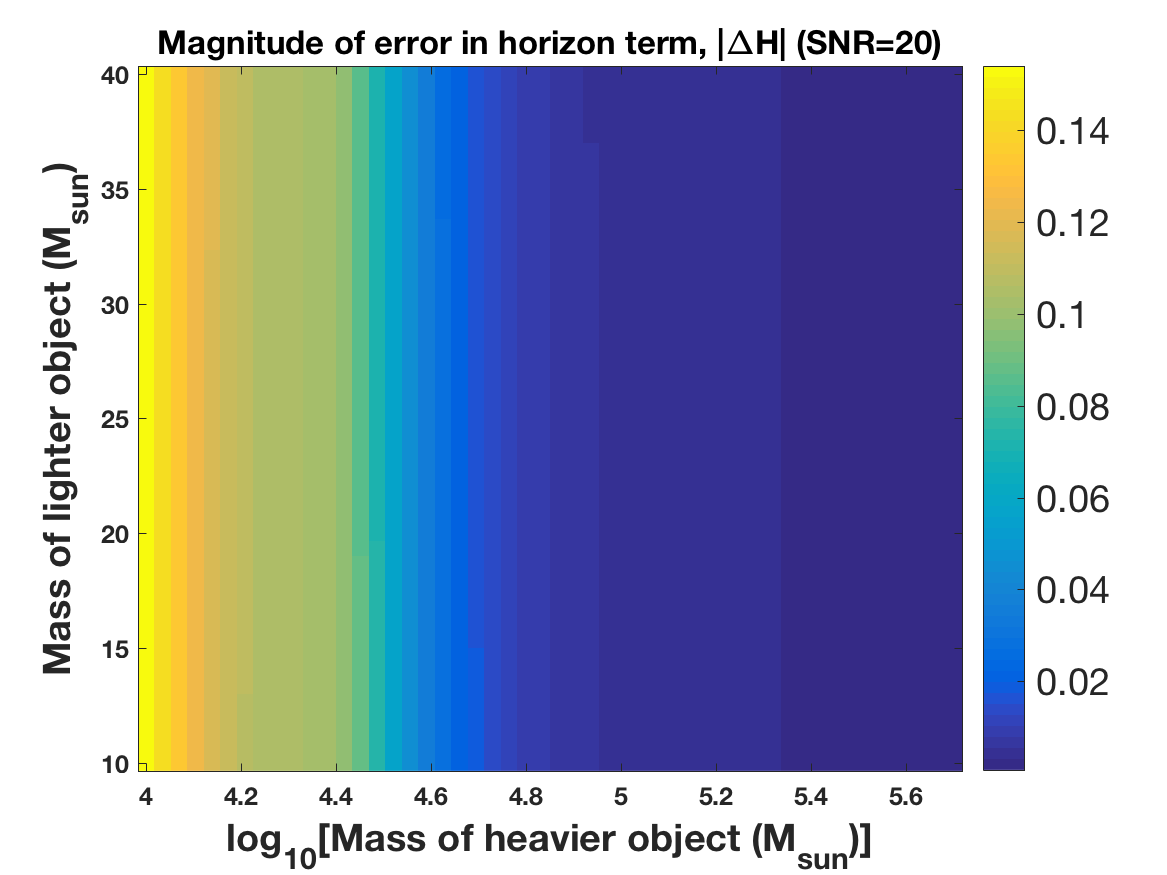}
\includegraphics[width=7.cm]{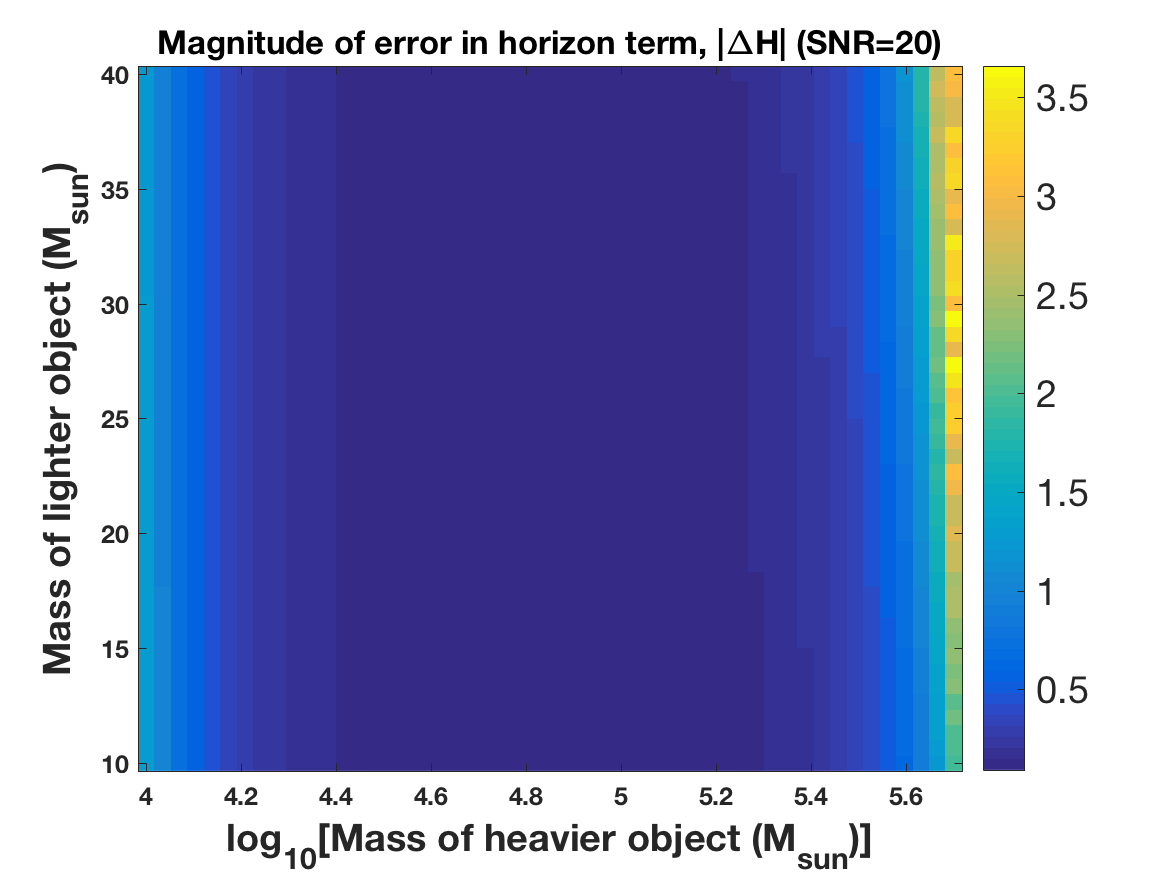}
\caption{Parameter estimation errors (in percentages, except for the bottom two plots) from a Fisher analysis are presented for EMRIs where both binary components are taken to be black holes. The spin of the supermassive black hole is $\chi_{BH}=0.9$ and that of the smaller companion is zero. In the left set of plots, it is assumed that the central object is a black hole and, hence, its mass and spin moments are all taken to be completely determined by its mass and spin. As shown above, in such a case, the errors in the estimation of the central object's mass, mass-ratio, and spin are quite small for a signal with an SNR of 20. We also consider a case in the bottom-left plot where the horizon parameter $H$ for the aforementioned system is taken to be unknown. In such a case, the horizon term can be determined to within a few percent of unity for central objects with mass $\gtrsim 2.5\times 10^4~M_{\odot}$. In the right set of plots errors in $\alpha_2$, $\alpha_3$, $\chi_{BH}$ and $H$ are shown for the same BBH EMRIs, except that for the measurement problem $\alpha_2$ and $\alpha_3$ are taken to be parameters independent of the mass and spin of the central object. Unsurprisingly, the inclusion of these two parameters among unknowns increases the errors for all parameters. The horizon term (bottom-right plot) is most adversely affected. Still, there are wide ranges of the central object mass value for which the errors are a few to several percent. Even the errors in the total mass and mass-ratio (not shown) are within a few percent.}
\label{FisherBBHChi0p9}
\end{figure*}

\end{widetext}

\begin{widetext}
\begin{figure*}
   \includegraphics[width=7.cm]{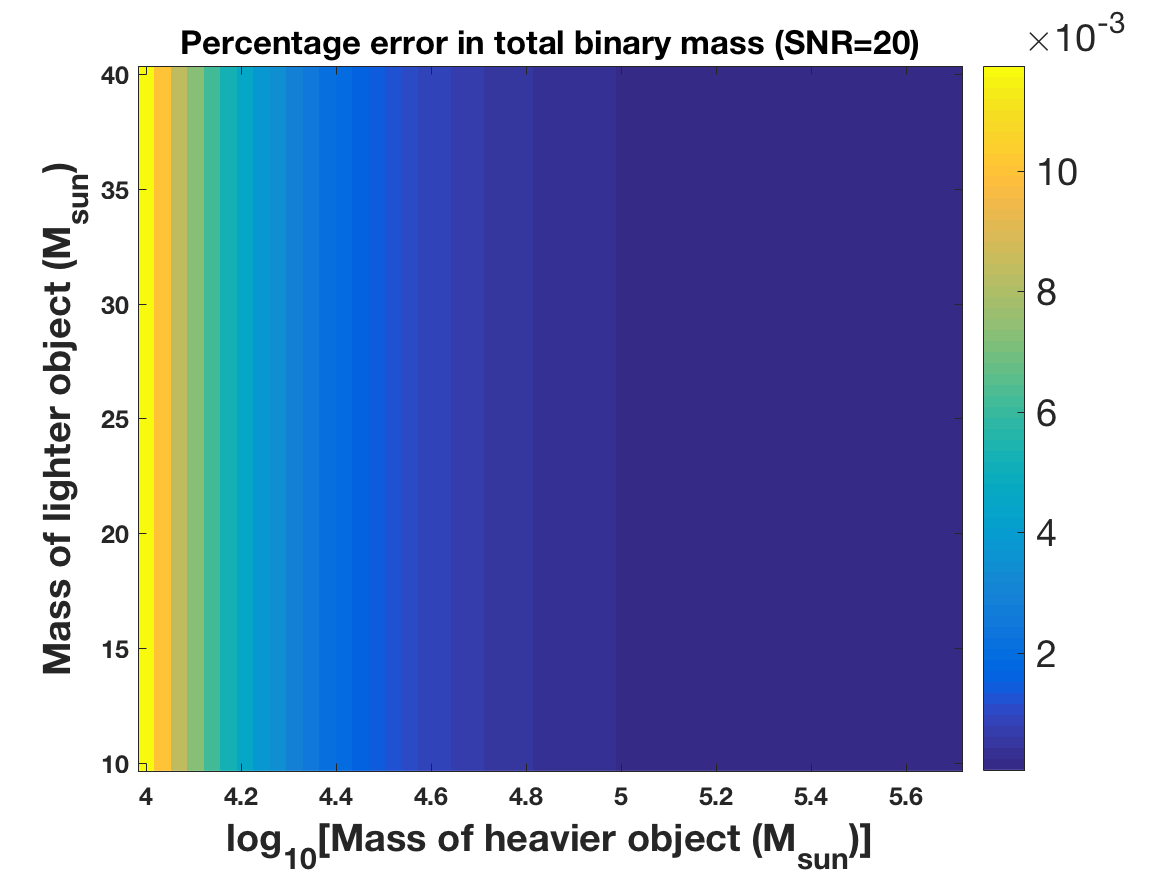}
\includegraphics[width=7.cm]{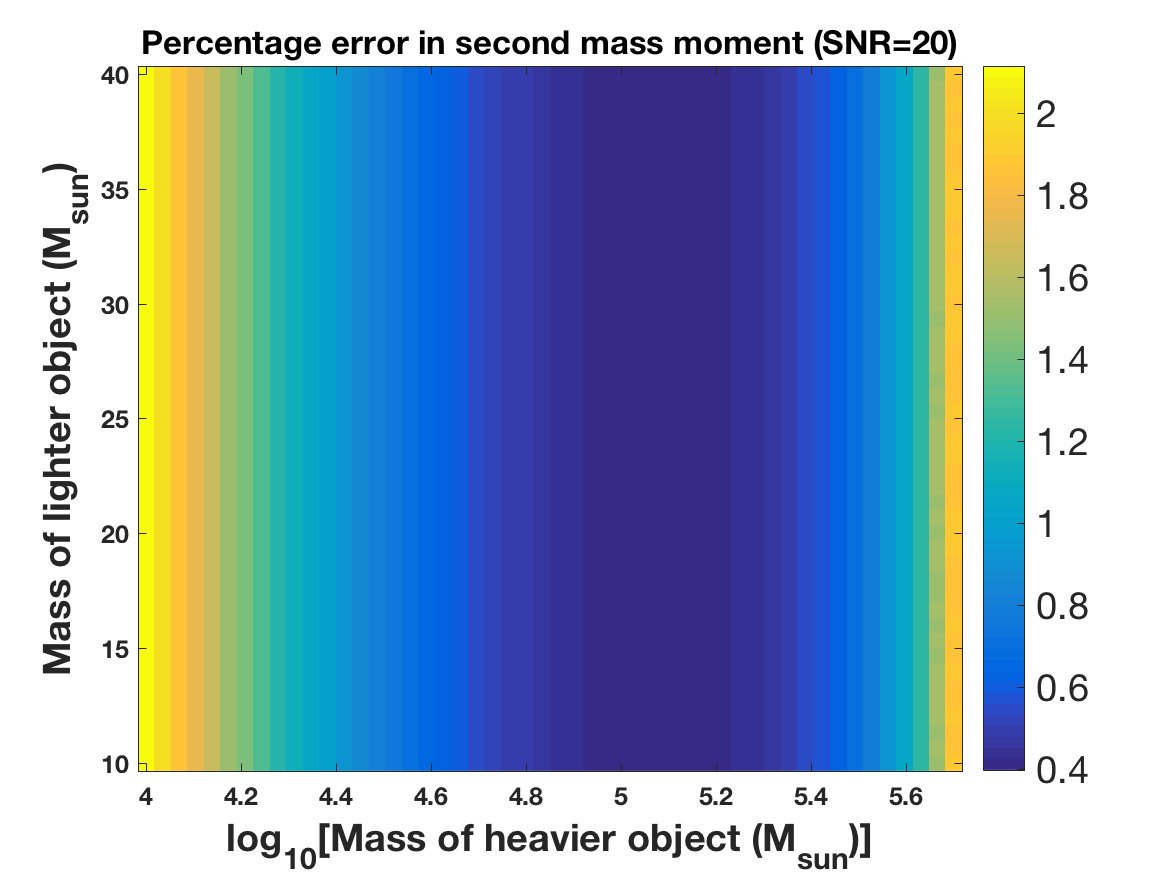}
\includegraphics[width=7.cm]{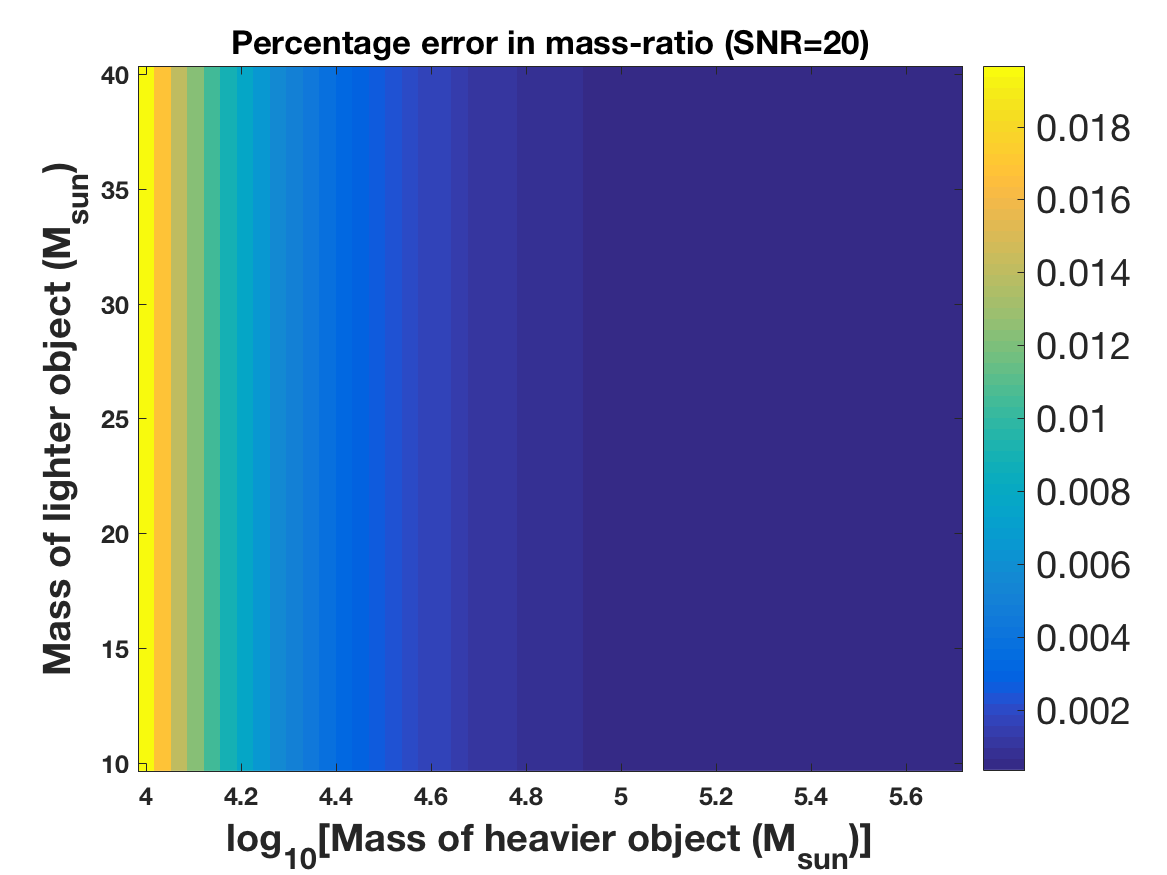}
\includegraphics[width=7.cm]{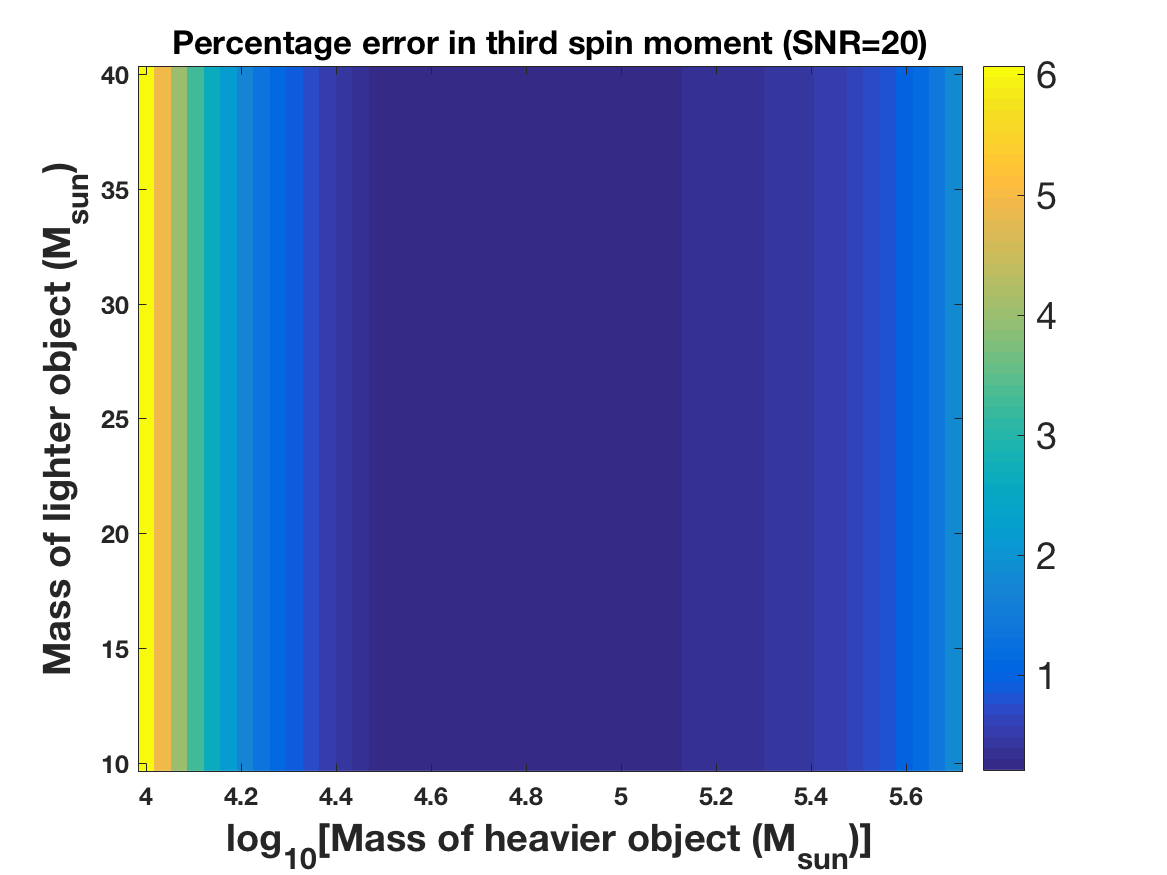}
\includegraphics[width=7.cm]{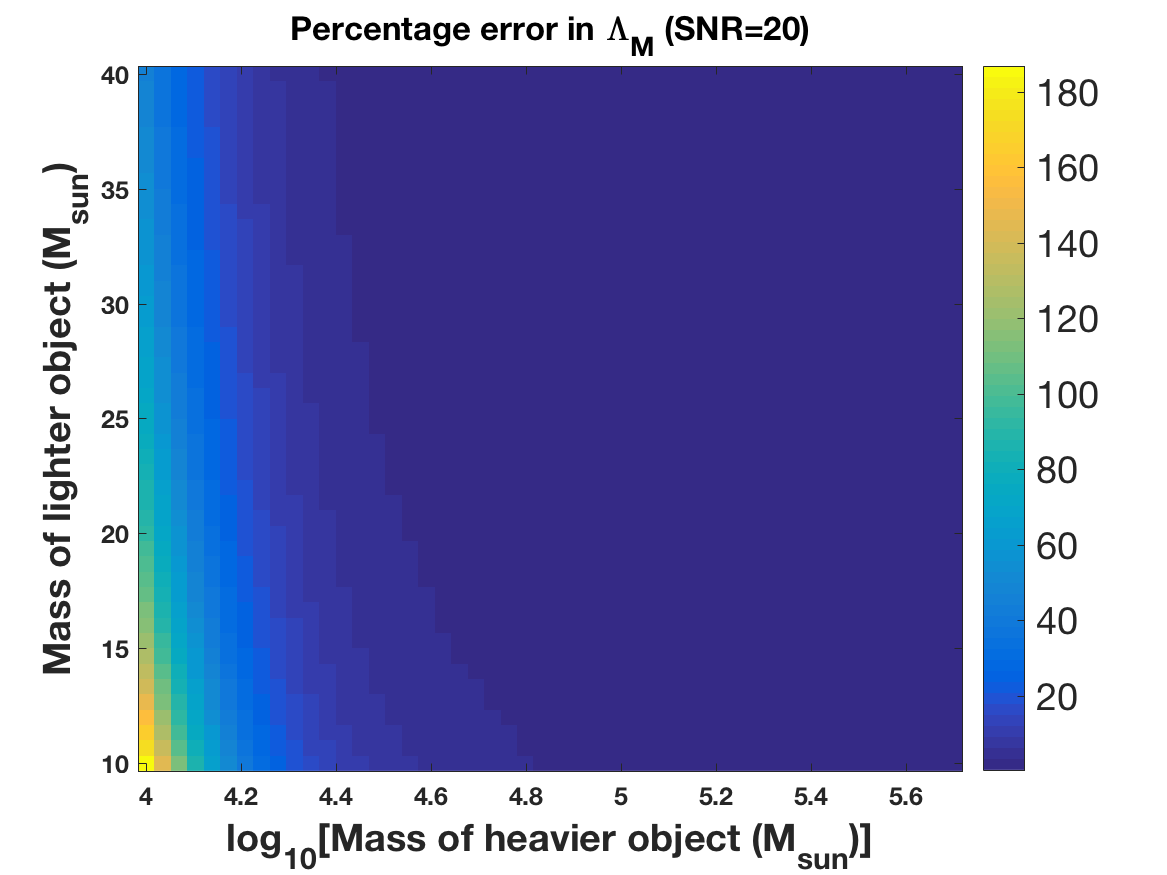}
\includegraphics[width=7.cm]{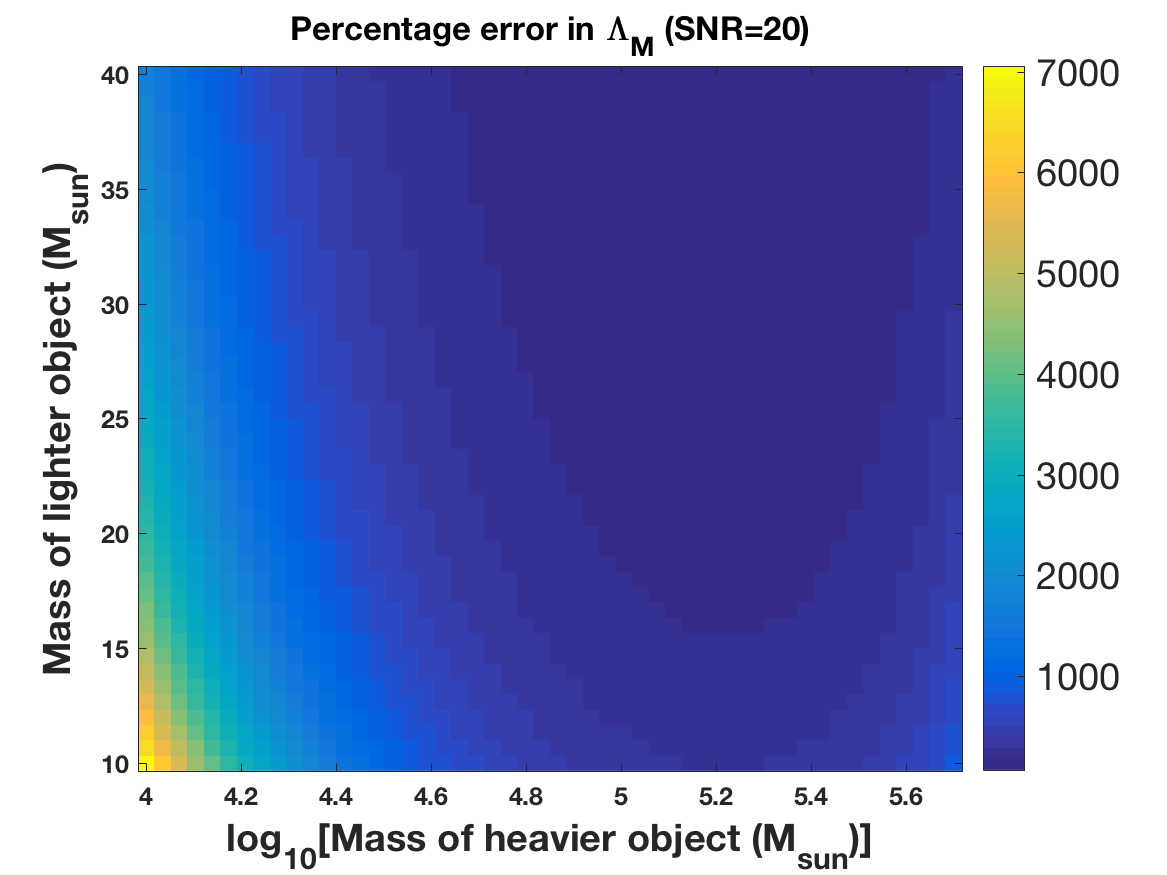}
\includegraphics[width=7.cm]{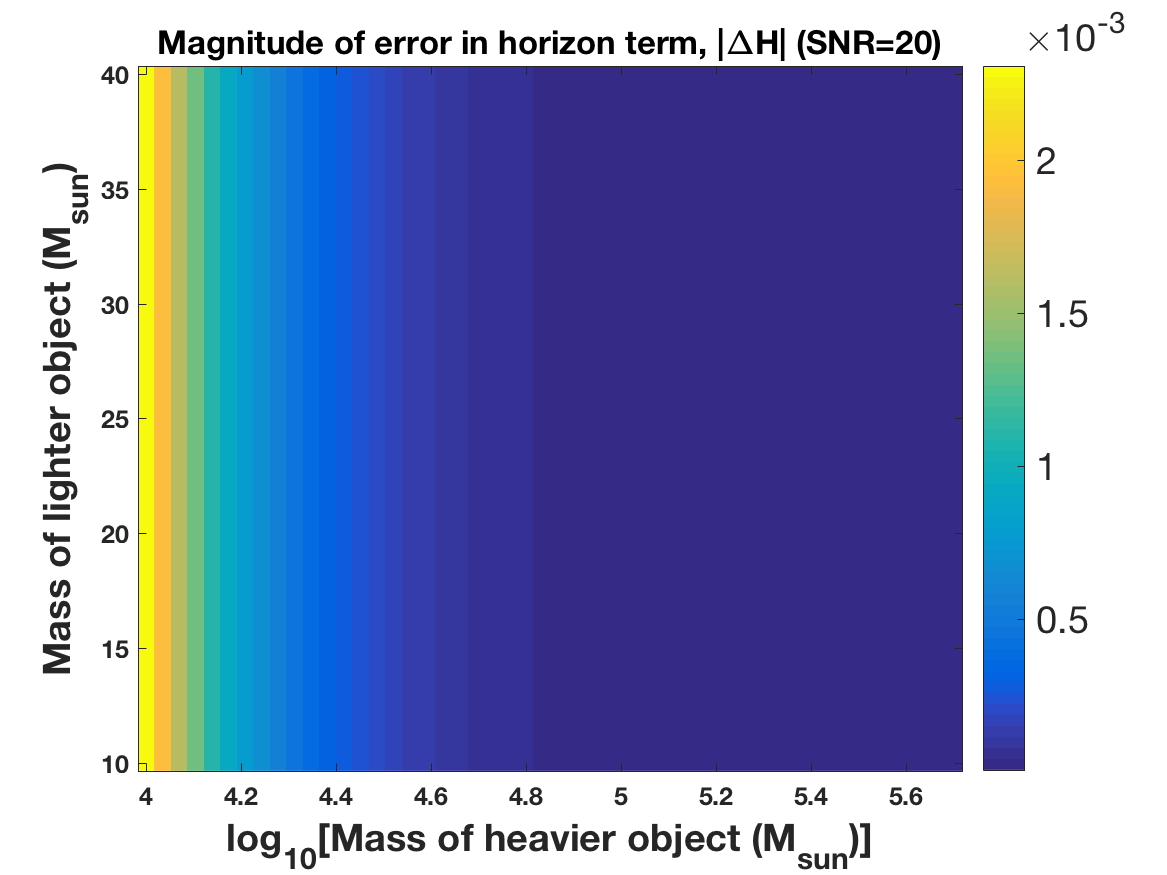}
\includegraphics[width=7.cm]{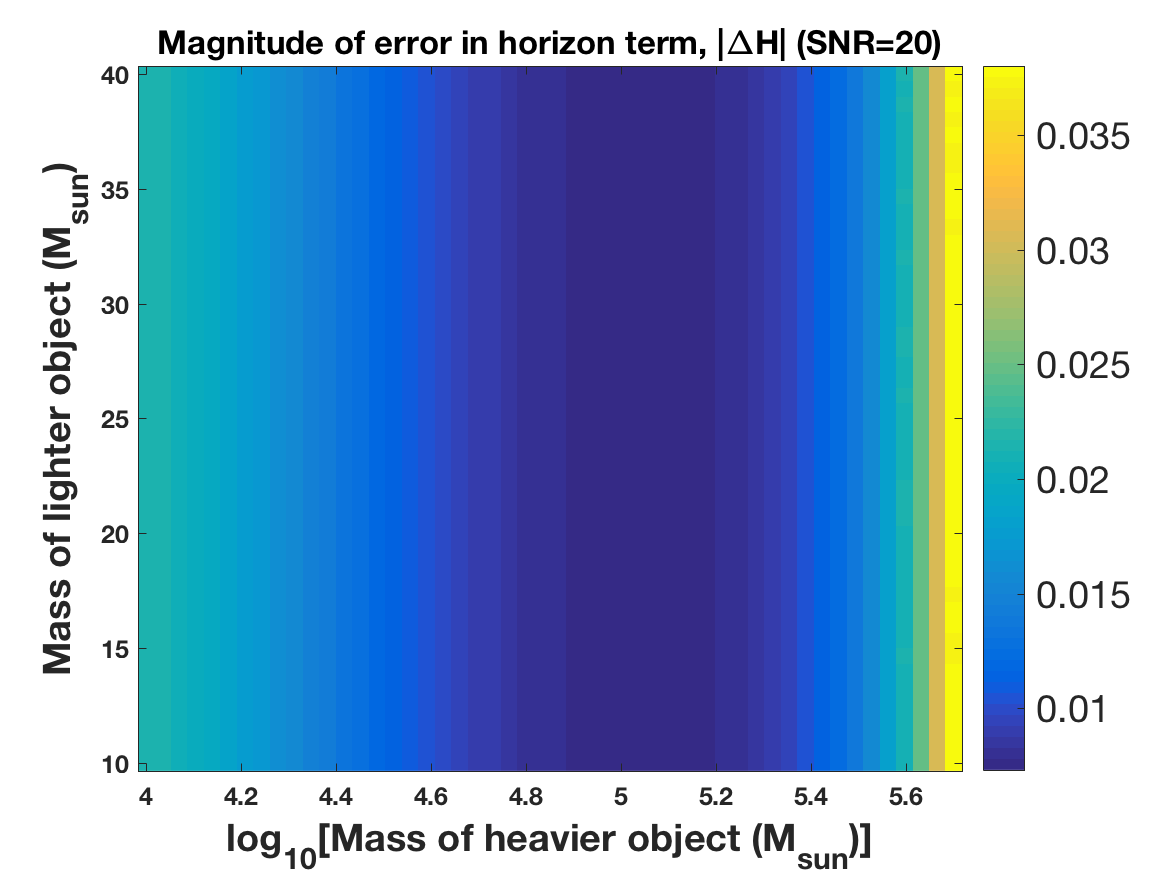}
\caption{Parameter estimation errors (in percentages, except for the bottom two plots) from a Fisher analysis are presented for EMRIs where the more massive component is a boson star (with $H=0$) whereas the lighter one is a black hole. The spin of the central object is $\chi=0.9$ and that of the smaller companion is zero. In the left set of plots the spin and the dimensionless tidal deformability parameter $\Lambda_M$ of the central object and the component masses are being measured. As shown above, in such a case, the errors in the estimation of the total mass, mass-ratio, and spin are quite small for a signal with an SNR of 20. In the right set of plots errors in $\alpha_2$, $\alpha_3$, $\chi$, $\Lambda_M$  and $H$ are shown for the same EMRIs. Here too the erros are within a few percent for the most part, except for $\Lambda_M$, which suffers large measurement erros. This suggests that the ability to measure $\Lambda_M$ is adversely affected by the absence of any prior knowledge of $\alpha_2$ and $\alpha_3$. Nevertheless, constraining $H$ to be close to zero in these cases is very much possible. 
}
\label{FisherBosonChi0p9}
\end{figure*}
\end{widetext}

We consider two kinds of central objects below, namely, supermassive black holes and supermassive boson stars. The lighter companion is always taken to be a black hole here.
When the central object is a black hole, its spin $(\chi)$ is defined in terms of the Kerr rotation parameter $a =M\chi$. In the case of a boson star (BS), its spin is defined in terms of its first spin moment $S_1$, namely, $\chi = S_1/M^2$. It is well known that the multipole moments of a Kerr BH are completely determined by its mass and spin as
\begin{equation}\label{Multipole expression for Kerr}
M_l + iS_l = M(ia)^l.
\end{equation}
In the case of a boson star this relation gets modified to~\cite{Ryan 1997, Berti 2006}
\begin{equation}\label{Multipole expression for BS}
M_l + iS_l = \alpha_lM(ia)^l\,,
\end{equation}
where $\alpha_l$ depends on $\chi$ and $M\mu_B^2/\sqrt[]{\lambda_{\rm boson}}$, with $\lambda_{\rm boson}$ and $\mu_B$ being the interaction strength of the quartic potential and the mass of the boson field, respectively, in  the massive boson star model~\cite{Ryan 1997}.~\footnote{ A slightly different approach can be found in Ref \cite{Krishnendu 2017, Krishnendu 2018}.} From the definition of the mass and spin, it follows that $\alpha_0 = \alpha_1 = 1$. A nice discussion regarding the three-hair relation for boson stars can be found in Ref.~\cite{Berti 2006}.  

In the Fisher analysis we used Eq.~(\ref{Multipole expression for BS}) for the multipole moments in the expression of signal phase for all EMRIs. When the central object is a black hole, we take $\alpha_l = 1$, for all $l$. Values of the $\alpha_l$ for boson stars are taken from Ref.~\cite{Ryan 1997}. We were unable to do a similar analysis for solitonic BS since multipole moments for such systems are not available in the literature. All the boson stars considered in the current work are massive BS. 

In Figs.~\ref{FisherBBHChi0p9} and \ref{FisherBosonChi0p9} by second mass moment and third spin moment we mean $\alpha_2$ and $\alpha_3$, respectively. For computing the errors presented in those figures, we set $\alpha_4 = 0 = \alpha_5$. This is because we found that the errors in $\alpha_4$ and $\alpha_5$ are quite high (i.e., mostly more than 100\% for signal-to-noise ratio (SNR) of 20). This implies that the signal is not very sensitive to variations in the values of these parameters. Whether more accurate waveform models will allow their determination at similar SNRs can be explored in the future.

In tables~\ref{BH_tab:errors} and \ref{BS_tab:errors}, we highlight parameter errors for some EMRIs not all of which are shown in Figs.~\ref{FisherBBHChi0p9} and \ref{FisherBosonChi0p9}. (Specifically, the $\chi = 0.5$ systems are not shown in the figures.) We notice that the errors generally reduce substantially with increasing spin. Since the spins of supermassive BHs may be quite high~\cite{Reynolds 2013}, we can expect the errors in the parameters of such systems to be smaller based on our analysis.

Broadly, we study the measurement precision of parameters for two kinds of system, namely, one binary where the central object is a super-massive black hole and another binary where that object is a boson star. We always take the smaller companion to be a black hole. While our formalism allows for non-black hole companions, we limit our scope here to the aforementioned systems for ease of interpreting and communicating our results. We use the LISA noise curve given in Ref.~\cite{Cornish 2017}. We have not accounted for the source confusion noise that is expected at frequencies a few times below 1~mHz to several times above that frequency, and will affect the parameter estimates of high-mass sources studied here.

\subsubsection{Central object as a black hole}

For our error analysis, the distance of the source is normalized such that the signal SNR remains fixed at 20. Moreover, we integrated all signals for the duration of 1 year. This means that when the total mass is small, most of the signal lies at higher frequencies where LISA sensitivity starts deteriorating. This is why the error increases as one reduces the mass of the heavier (or central) object, which dominates the contribution to the total mass. For heavier EMRIs too the parameter errors can worsen because most of the signal lies at frequencies below the most sensitive part of the LISA band, which is around 8-9~mHz. 
This aspect of the error distribution should be revisited for more accurate signal models. 

In Fig.~\ref{FisherBBHChi0p9} parameter estimation errors (in percentages) from a Fisher analysis are presented for EMRIs where both binary components are taken to be black holes. The spin of the supermassive black hole is $\chi_{\rm BH}=0.9$ and that of the smaller companion is zero. In the left set of plots, the mass and spin moments are all taken to be completely determined by the  mass and spin of the central object as if it were a black hole. In such a case, the errors in the estimation of the central object's mass, mass-ratio, and spin are quite small for a signal with an SNR of 20. Our errors for this case are consistent with those presented in Ref.~\cite{Gair 2017}, even if on the higher side. Our larger errors can be attributed partly to our different waveform model but mostly to the fact that we are estimating a larger number of parameters here. We also consider a case in the bottom-left plot of Fig.~\ref{FisherBBHChi0p9} where the horizon parameter $H$ for the aforementioned system is taken to be unknown. In such a case, the horizon term can be determined to within a few percent of unity for central objects with mass $\gtrsim 2.5\times 10^4~M_\odot$. In the right set of plots of Fig.~\ref{FisherBBHChi0p9}, the errors in $\alpha_2$, $\alpha_3$, $\chi_{BH}$ and $H$ are shown for the same BBH EMRIs, except that for the measurement problem $\alpha_2$ and $\alpha_3$ are taken to be parameters independent of the mass and spin of the central object. Unsurprisingly, the inclusion of these two parameters among unknowns increases the errors for all parameters. The horizon term (bottom-right plot) is most adversely affected. Still, there are wide ranges of the central object mass value for which the errors are a few to several percent. Even the errors in the total mass and mass-ratio (not shown) are within a few percent. 

We also computed the errors for the case where the spin of the central object is smaller -- at $\chi= 0.5$. 
Table~\ref{BH_tab:errors} (table~\ref{BS_tab:errors}) compares them with the errors for the $\chi = 0.9$ case when the central object is a black hole (boson star) and we are only measuring its spin, the binary's total mass, mass-ratio, and $H$ (as well as $\Lambda_M$ for the BS). As seen there, the errors tend to increase when $\chi$ decreases from 0.9 to 0.5.

{\centering
\begin{table}[h]
\begin{tabular}{| p{1.6cm} | p{0.6cm} | p{0.8cm}| p{0.9cm}| p{1.cm}  | p{0.9cm}| }
    \hline
    ${M_{\rm Cen}}~(M_\odot)$ & $\chi_{\rm BH}$ & $\frac{\Delta\chi_{\rm BH}}{\chi_{\rm BH}}$ & $|\Delta H|$ & $\frac{\Delta  M_{\rm tot}}{M_{\rm tot}}$ & $\frac{\Delta q}{q}$ \\ \hline
    $10^5$& .9&.005 &.01 &.0025 &.004 \\ \hline
    4$\times 10^4$&.9 &.02 &.04 &.005 &.01 \\ \hline
    $10^5$& .5& 0.01&.025 &.003 &.005 \\ \hline
    $4\times 10^4$& .5& 0.08&.28 &.025 & .04\\ \hline
\end{tabular}
\caption {Effect of central object's spin on parameter errors: A selection of parameter errors (in percentage, except for $H$) from the plots  in the left column in Fig.~\ref{FisherBBHChi0p9} are listed in the last four columns for a black hole as the central object, with spin of 0.9. Additionally, for comparison, we present errors when the central object spin is 0.5.} 
\label{BH_tab:errors}
\end{table}
}

{\centering
\begin{table}[h]
\begin{tabular}{| p{1.6cm} | p{0.6cm} | p{0.8cm}| p{0.9cm}| p{1.4cm}  | p{1.4cm}| }
    \hline
    ${M_{\rm Cen}}~(M_\odot)$ & $\chi_{\rm BS}$ & $\frac{\Delta\Lambda_{ M}}{\Lambda_{M}}$ & $|\Delta H|$ & $\frac{\Delta  M_{\rm tot}}{M_{\rm tot}}$ & $\frac{\Delta q}{q}$ \\ \hline
    $10^5$& .9&10 &.0001 &$10^{-3}$ & $10^{-3}$\\ \hline
    4$\times 10^4$&.9 &15 &.00025 & $1.5\times 10^{-3}$&.002 \\ \hline
    $10^5$& .5& 10& .001& $5\times 10^{-3}$& $10^{-2}$\\ \hline
    $4\times 10^4$& .5& 20& .002& $10^{-2}$& $1.5\times 10^{-2}$\\ \hline
\end{tabular}
\caption {Effect of central object's spin on parameter errors: A selection of parameter errors (in percentage, except for $H$) from the plots  in the left column in Fig.~\ref{FisherBosonChi0p9} are listed in the last four columns for a boson star as the central object, with spin of 0.9. Additionally, for comparison, we present errors when the central object spin is 0.5.} 
\label{BS_tab:errors}
\end{table}
}

\subsubsection{Central object as a boson star}

Depending on the bare mass~$(\mu_{B})$ of the boson field and the nature of the interaction, the mass of a BS can take a range of values. These values can even be $> 10^6 M_{\odot}$~\cite{Mass of the Boson star}. For this study, 
the values of $\lambda_M / M^5~(\equiv\Lambda_M)$ have been taken from the work by Senett et al.~\cite{Tidal deformation of BS}. There $\Lambda_M$ has been expressed in terms of $\frac{M \mu_{B}}{m^2_p}$, where $m_p$ is the Planck mass. Therefore, for a given value of $M$, the value of $\Lambda_M$ depends on how light the boson field is. 

In Fig.~\ref{FisherBosonChi0p9} parameter estimation errors (in percentages) from a Fisher analysis are presented for EMRIs where the more massive component is a boson star (with $H=0$) whereas the lighter one is a black hole. The spin of the central object is $\chi=0.9$ and that of the smaller companion is zero. In the left set of plots the spin and the dimensionless tidal deformability parameter $\Lambda_M$ of the central object and the component masses are being measured. As shown there, in such a case, the errors in the estimation of the total mass, mass-ratio, and spin are quite small for a signal with an SNR of 20. In the right set of plots errors in $\alpha_2$, $\alpha_3$, $\chi$, $\Lambda_M$  and $H$ are shown for the same EMRIs. Here too the errors are within a few percent for the most part, except for $\Lambda_M$, which suffers large measurement errors. This suggests that the ability to measure $\Lambda_M$ is adversely affected by the absence of any prior knowledge of $\alpha_2$ and $\alpha_3$. Nevertheless, constraining $H$ to be close to zero in these cases remains a possibility.

\section{Discussion}
\label{sec:testing no hair}

It is already understood that the multipole moments of the central body provide information about its vacuum space time. Therefore, we can deduce from those moments the nature of the central object. Owing to that we can test the No-hair theorem, and check whether the black hole uniqueness theorem holds or not~\cite{no hair israel,no hair israel2,no hair wald,no hair carter,no hair Robinson}. 

To test the No-hair theorem we need two pieces of information. One of them is whether the central object has a horizon (i.e., is a black hole), with the value of $H$ observationally consistent with unity. The other one is the knowledge of the multipole moments, from the observed GW emission; these moments will reveal if the central object has any hairs. 

 From Fig.~\ref{FisherBBHChi0p9}  we notice that the error in $H$ is less than $50\%$ if the mass of the central BH is $\leq 5\times 10^{5}~\rm{M_{\odot}}$. This implies that for such a case the value of $H$ can be determined more precisely than $\sim 1\pm.5$ (at the 1$\sigma$ level). (If the SNR is 50~\cite{Gair 2017}, this error reduces to 30\%, which is a possibility.) When the central object is a BS the situation is much better: From Fig.~\ref{FisherBosonChi0p9} we infer that the error in $H$ is less than $4\%$ for the entire range of masses of the BS. Therefore, the value of $H$ can be determined to be more precisely than $0.00\pm.04$. This suggests that these two systems can be distinguished from each other, at the 1$\sigma$ level. For this reason, it is important that one revisits this estimation problem with more accurate waveform models.


Owing to the aforementioned results, testing the No-hair theorem in EMRIs remains a viable pursuit. In the figures here we have shown how precisely the first few mass and spin moments are measurable. From Fig.~\ref{FisherBBHChi0p9} we notice that for central BH masses greater than $10^{4}~ \rm{M_{\odot}}$, $\alpha_2$ and $\alpha_3$ can be measured with better precision than $1.0\pm 0.4$ and $1.0\pm0.8$, respectively. These errors reduce when the central object is a BS (for the same SNR). The injected value of $\alpha_2$ and $\alpha_3$ are 34 and 47, respectively. From Fig.~\ref{FisherBosonChi0p9} we notice that for the entire mass range of BS, $\alpha_2$ and $\alpha_3$ can be measured more precisely than $34.00\pm 0.68 $ and $47.00 \pm 2.82$, respectively. With an accurate measurement of $H$ we will be able to distinguish between black holes and boson stars as central objects in EMRIs. This implies that it is likely that the No-hair theorem for BHs will be testable by measuring the multipole moments with required precision.


\section*{Acknowledgments}

We would like to thank Sanjeev Dhurandhar and Geoffrey Lovelace for helpful discussions. This work is supported in part by the Navajbai Ratan Tata Trust and NSF grant PHY-1506497. SD would like to thank University Grants Commission (UGC), India, for financial support as senior research fellow.

\appendix
\section{Expressions\label{expressions}}

It was discussed in Eq.~(\ref{N series}) that $\Delta N$ can be expressed as follows,
\begin{equation}
\Delta N = \frac{5 }{96 \pi  q v^5}\sum_{n =0}^{10}N_n v^n,
\end{equation}
The expressions for $N_n$ are listed below:

\begin{widetext}
\begin{equation}
\begin{split}
N_0 =& 1\\
N_2 =& \frac{743 }{336}\\
N_3 =&  \bigg(\frac{113 S_1}{12 M^2}-4 \pi
   \bigg)\\
   N_4 =&  \bigg(-\frac{S_1^2}{16 M^4}+\frac{5 M_2}{M^3}+\frac{3058673}{1016064}\bigg)\\
       N_5 =&  \bigg(-A_1+\frac{150323 S_1}{2016 M^2}-\frac{995 \pi }{42}\bigg)\\
         N_6 =&  \bigg(-\beta +\frac{-130816 \pi  M^2 S_1+135640 M_2 M+84741 S_1^2}{2688 M^4}+\frac{1712 \log (v)}{105}+16 \pi
   ^2-\frac{6867871393}{341397504}\bigg)\\
    N_7 =&  \bigg(-\frac{995 A_1}{168}-B_1-\frac{65 S_1^3}{96 M^6}+\frac{\pi  S_1^2}{2
   M^4}-\frac{22 S_3}{M^4}+\frac{5823667355 S_1}{12192768 M^2}+M_2 \bigg(\frac{557 S_1}{12 M^5}-\frac{28 \pi }{M^3}\bigg)-\frac{30400075 \pi }{254016}\bigg)\\
   N_8 =&  \bigg(A_1 \bigg\{8 \pi -\frac{73 S_1}{6
   M^2}\bigg\}-\frac{995 \beta }{168}-C_1+\frac{1}{16257024
   M^8}\bigg\{-9204217344 \pi  M^6 S_1+\bigg(6132787781 M^3-6096384 M_2\bigg) M S_1^2\\
   &+48 \bigg(119672783 M_2 M^3-3810240 M_4 M+5842368 M_2^2\bigg) M^2+63504 S_1^4\bigg\}+\frac{9203 \log (v)}{210}+\frac{1079 \pi ^2}{7}-\frac{221821577506343}{1032386052096}\bigg)\\
    N_9 =&  \bigg(A_1 \bigg(\frac{S_1^2-56 M M_2}{8 M^4}-\frac{30400075}{1016064}\bigg)+\pi  \bigg(8 \beta -64 \pi ^2-\frac{3417184151}{7112448}\bigg)-\frac{995
   B_1}{168}-D_1-\frac{18155 S_3}{84 M^4}\\
   &+\frac{M_2 \bigg(1232627 S_1-745392 \pi  M^2\bigg)}{2016 M^5}+\frac{S_1}{11266117632 M^6} \bigg(\bigg\{-137071097856 \beta +2688846741504 \pi ^2\\
   &+29876928310087\bigg\} M^4-38808 S_1 \bigg(71137296
   \pi  M^2-20596235 S_1\bigg)\bigg)+\frac{62488}{315} S_1 \chi  \log (v)-\frac{6848}{105} \pi  \log (v)\bigg)\\
   N_{10} =& \bigg(A_1 \bigg(\frac{1079 \pi }{14}-\frac{285349 S_1}{2016 M^2}\bigg)+A_1^2-A-\frac{30400075 \beta }{1016064}+B_1 \bigg(8 \pi -\frac{73 S_1}{6 M^2}\bigg)-\frac{995 C_1}{168}-F_1-\frac{82585 S_1^4}{28672
   M^8}+\frac{836035 M_2 S_1^2}{4032 M^7}\\
   &+\frac{163 \pi  S_1^3}{24 M^6}-\frac{971 S_3 S_1}{6 M^6}+\frac{165929 M_2^2}{672 M^6}-\frac{1030 \pi  M_2 S_1}{3 M^5}-\frac{21289 M_4}{168 M^5}+\frac{\beta  S_1^2}{8
   M^4}+\frac{4528045764947 S_1^2}{1365590016 M^4}-\frac{3 \pi ^2 S_1^2}{M^4}+\frac{104 \pi  S_3}{M^4}\\
   &-\frac{7 \beta  M_2}{M^3}+\frac{7750789809073 M_2}{3755372544 M^3}+\frac{144 \pi ^2 M_2}{M^3}-\frac{3628680619 \pi 
   S_1}{762048 M^2}+\frac{3 X}{M q}+\frac{1}{110020680}\bigg\{-12557026944 M_2 \chi ^2 \log (v)\\
   &-224232624 S_1^2 \chi ^2 \log (v)+6470582647 \log (v)\bigg\}+\frac{11956093 \pi
   ^2}{10584}-\frac{165899152309973251}{115627237834752}\bigg)\\
     \end{split}
\end{equation}
\end{widetext}

 The expression for phase $\psi(f)$ is given in Eq.~(\ref{phase}),
\begin{equation}
\begin{split}
\psi(f) = 2\pi f t_c - 2\phi_c-\frac{\pi}{4} +  I(v) -  I(v_i)
\end{split}
\end{equation}

\begin{equation}
I(v) - I(v_i) = \int^v_{v_i} d\bar{v} (v^3-\bar{v}^3) \frac{6\pi\Delta N}{\bar{v}^4}
\end{equation}

\begin{equation}
I(v) = \sum_{n=-5}^{5}I_n (v) v^n,
\end{equation}
The expression for $I_n(v)$ are given below:
\begin{widetext}
\begin{equation}
\begin{split}
I_5 = &\bigg(-\frac{247755 S_1^4}{917504 M^8 q}+\frac{163 \pi  S_1^3}{256 M^6 q}+\frac{107 \chi ^2 S_1^2}{800 q}+\frac{3 \beta  S_1^2}{256 M^4 q}-\frac{107 \chi ^2 \log (v) S_1^2}{560 q}+\frac{836035 M_2 S_1^2}{43008 M^7
   q}+\frac{4528045764947 S_1^2}{14566293504 M^4 q}-\frac{9 \pi ^2 S_1^2}{32 M^4 q}\\
   &-\frac{515 \pi  M_2 S_1}{16 M^5 q}-\frac{3628680619 \pi  S_1}{8128512 M^2 q}+\frac{3 A_1^2}{32 q}+\frac{165929 M_2^2}{7168 M^6
   q}+\frac{9 X}{32 M q^2}+\frac{-\frac{3 A}{32}-\frac{30400075 \beta }{10838016}+\frac{11956093 \pi ^2}{112896}-\frac{46931328079382799929}{339173230981939200}}{q}\\
   &+\frac{6470582647 \log (v)}{1173553920 q}-\frac{995
   C_1}{1792 q}-\frac{3 F_1}{32 q}+\frac{749 \chi ^2 M_2}{100 q}-\frac{21 \beta  M_2}{32 M^3 q}-\frac{107 \chi ^2 \log (v) M_2}{10 q}+\frac{7750789809073 M_2}{40057307136 M^3 q}+\frac{27 \pi ^2 M_2}{2 M^3
   q}\\
   &-\frac{21289 M_4}{1792 M^5 q}+\frac{A_1 \bigg(155376 M^2 \pi -285349 S_1\bigg)}{21504 M^2 q}+\frac{B_1 \bigg(48 M^2 \pi -73 S_1\bigg)}{64 M^2 q}+\frac{\bigg(624 M^2 \pi -971 S_1\bigg) S_3}{64 M^6 q}\bigg)\\
   I_4 = &-\frac{1}{240343842816 M^6 q}\bigg(1584 \pi  \bigg(-284497920 \beta +2319335424 \log (v)+2275983360 \pi ^2+14186751475\bigg) M^6\\
   &+333624614400 B_1 M^6+56330588160 D_1 M^6+13968197591040 \chi  S_1 M^6-11174558072832 \chi  \log (v) S_1
   M^6\\
   &+685355489280 \beta  S_1 M^4-13444233707520 \pi ^2 S_1 M^4-149384641550435 S_1 M^4+20827564369920 \pi  M_2 M^3\\
   &+13803480915840 \pi  S_1^2 M^2+55440 A_1 \bigg(30400075 M^4+7112448 M_2 M-127008 S_1^2\bigg)
   M^2+12174783667200 S_3 M^2\\
   &-34441767803520 M_2 S_1 M-3996493439400 S_1^3\bigg) \\
   I_3 = &\frac{1}{49554530500608 M^8 q}\bigg(\bigg(30572146483200 \beta +3 \log (v) \bigg(-30572146483200 \beta +113107724255232 \log
   (v)\\
   &+795674678722560 \pi ^2-1184513037035203\bigg)-795674678722560 \pi ^2+1184513037035203\bigg) M^8\\
   &+317520 (3 \log (v)-1) \bigg(-32256 \bigg(6132 A_1+285349 \pi \bigg) S_1 M^6+48 \bigg(2709504 \pi  A_1 M^6-338688
   C_1 M^6\\
   &+119672783 M_2 M^3-3810240 M_4 M+5842368 M_2^2\bigg) M^2+\bigg(6132787781 M^3-6096384 M_2\bigg) S_1^2 M+63504 S_1^4\bigg)\bigg)\\
   I_2 = &\frac{5}{130056192 M^6 q}\bigg(72213120 A_1 M^6+12192768 B_1
   M^6+1459203600 \pi  M^6-5823667355 S_1 M^4+341397504 \pi  M_2 M^3\\
   &-6096384 \pi  S_1^2 M^2+268240896 S_3 M^2-565947648 M_2 S_1 M+8255520 S_1^3\bigg)\\
   I_1 = &\frac{ 1}{3641573376 M^4 q}\bigg(\bigg(1706987520 \beta
   -27832025088 \log (v)-27311800320 \pi ^2+48255369509\bigg) M^4+635040 \bigg(\bigg(130816 M^2 \pi \\
   &-84741 S_1\bigg) S_1-135640 M M_2\bigg)\bigg)\\
   I_0 = &\frac{5 (3 \log (v)+1) \left(48 M^2 \left(42
   A_1+995 \pi \right)-150323 S_1\right)}{96768 M^2 q}\\
   I_{-1} = &\frac{5 \left(3058673 M^4+5080320 M_2 M-63504 S_1^2\right)}{21676032 M^4 q }\\
   I_{-2} = &\frac{\frac{113 S_1}{M^2}-48 \pi }{128 q }\\
   I_{-3} = &\frac{3715}{32256 q }\\
   I_{-5} = &\frac{3}{128 q   }.
\end{split}
\end{equation}
\end{widetext}

\end{document}